\documentclass[12pt,letterpaper]{article}

\usepackage[left=1in,right=1in,top=1in,bottom=1in]{geometry}
\usepackage{setspace}
\usepackage{amsmath,amssymb,amsthm,mathtools,bm}

\usepackage{graphicx}
\usepackage{booktabs}
\usepackage{threeparttable}
\usepackage{longtable}
\usepackage{array}
\usepackage{multirow}
\usepackage{caption}
\usepackage{subcaption}
\usepackage{float}
\usepackage{placeins}
\usepackage{makecell}

\usepackage{enumitem}
\usepackage{xcolor}
\usepackage{appendix}
\usepackage{pdflscape}
\usepackage{csquotes}

\usepackage{chngcntr}
\counterwithin{figure}{section}
\counterwithin{table}{section}

\usepackage[authoryear]{natbib}

\usepackage[colorlinks=true,linkcolor=blue,citecolor=blue,urlcolor=blue]{hyperref}
\usepackage[nameinlink,capitalize]{cleveref}

\theoremstyle{definition}

\title{The Cost of a Free Lunch: \\Evidence from U.S. Derivatives Markets}
\author{Useong Shin\thanks{
		Sogang Business School, Sogang University (Seoul, Korea).\\
		ORCID: \href{https://orcid.org/0009-0003-0197-9003}{0009-0003-0197-9003}\\
		Email: \texttt{useong@sogang.ac.kr}
}}
\date{\today}

\begin{document}
	
	\maketitle
	\thispagestyle{empty}
	
	\begin{flushleft}
	\end{flushleft}
	
	\noindent\textbf{Acknowledgments:}
	I am grateful to Michele Azzone (Politecnico di Milano) for generously sharing OIS data, for guidance on implementing the implied-discount-factor pipeline, and for detailed feedback on earlier drafts; to Baeho Kim (Korea University) for helpful discussions on the theoretical landscape of path-risk pricing and limits to arbitrage; and to Chaehwan Won (Sogang University) for raising critical questions about potential measurement artifacts that helped shape the robustness design of the draft. All remaining errors are my own.
	
	\begin{abstract}
		Put--call parity is a terminal-payoff identity;
		quoted residuals against traded futures are near zero.
		Yet enforcing parity is path-dependent,
		exposing arbitrageurs to daily settlement, margin, and finite capital.
		Using minute-level NBBO data on S\&P~500 and Russell~2000 options,
		I extract option-implied discount factors, compare them with the OIS curve,
		and construct an annualized \textit{carry gap}
		(sample median $\approx$37\,bp, $>$98\% positive).
		A reduced-form specification centered on a volatility$\times\!\sqrt{\tau}$
		path-risk term links the carry gap to implementation risk,
		trading frictions, and financial conditions,
		with coefficient signs stable across leave-one-year-out validation.
		The carry gap is an implementation wedge invisible in price space
		but systematic in carry space.
	\end{abstract}
	
	\newpage
	\pagenumbering{arabic}
	
	\section{Introduction}
	\label{sec:intro}
	
\begin{figure}[H]
		\centering
		\includegraphics[width=6.5in]{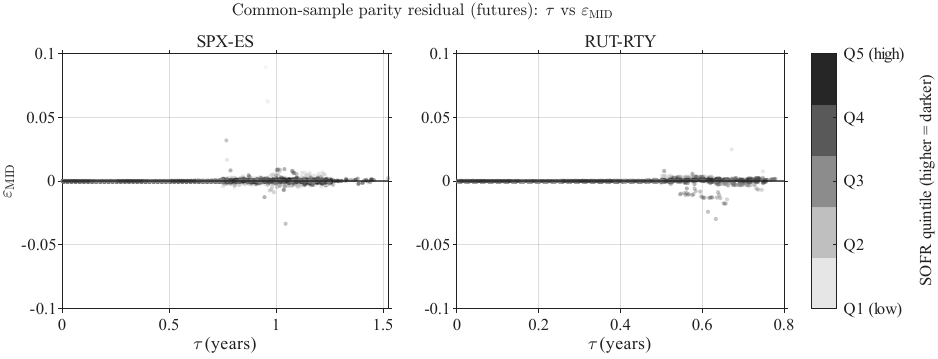}
		\caption{Quoted put--call parity residual computed against
			traded futures-implied forwards.
			Residuals cluster tightly around zero in both markets.}
		\label{fig:CP_resid_fut}
	\end{figure}

	Put--call parity is among the most fundamental no-arbitrage relations.
	Combining a European call, a put at the same strike and maturity, the
	underlying, and a risk-free bond locks in a deterministic terminal payoff.
	European index options provide an especially clean empirical setting for this
	relation. Early exercise is absent, exchange-traded futures provide a direct
	proxy for the forward, and short exposure can be implemented through futures
	without stock-borrow constraints. If any market should compress put--call
	parity residuals, it is the market for SPX and RUT index options. Consistent
	with this view, quoted parity residuals measured against traded futures are
	tightly compressed around zero, as Figure~\ref{fig:CP_resid_fut} shows.
	
	\FloatBarrier
	\FloatBarrier
	
\begin{figure}[H]
		\centering
		\includegraphics[width=6.5in]{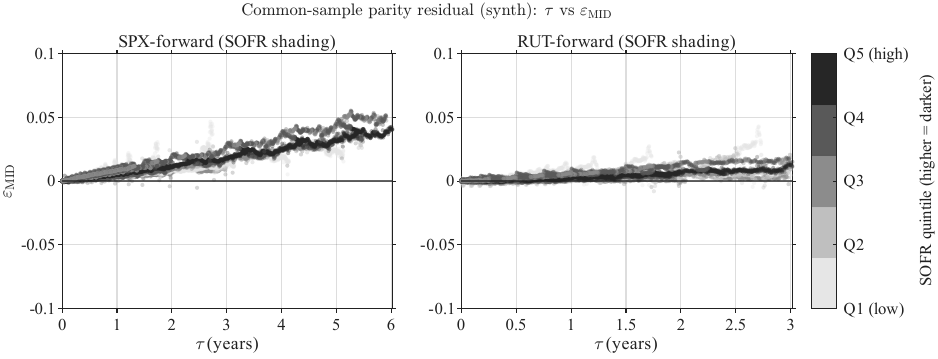}
		\caption{Parity residual implied by the option-cross-sectional
			synthetic-forward identification, benchmarked against OIS.
			Unlike the traded-futures benchmark, a systematic carry-space residual
			structure is visible.}
		\label{fig:CP_resid_synth}
	\end{figure}

	This clean benchmark also sharpens the puzzle. The absence of a visible
	price-space residual does not imply that enforcing parity is costless. When
	option-implied discount factors are extracted from the option cross-section
	and compared with an OIS benchmark, a different object becomes visible.
	Figure~\ref{fig:CP_resid_synth} shows that even when futures-based residuals
	remain small, a systematic residual structure appears on a carry basis.
	
	\FloatBarrier
	\FloatBarrier
	
	This contrast motivates the paper. Put--call parity contains two distinct
	claims. The first is a terminal-payoff identity. The second is the stronger
	claim that the strategy enforcing this identity is economically risk-free in
	practice. The former is exact by construction. The latter need not follow once
	daily settlement, variation margin, funding costs, nonsynchronous execution,
	illiquidity, and finite capital are taken seriously.
	
	Consider an arbitrageur, Bob, who observes a positive put--call parity residual
	and enters \(+C-P-F\): long call, short put, and short futures. At maturity,
	the option-implied forward and the futures leg offset, so the terminal payoff
	is deterministic and the residual is harvested. At inception, the position may
	require little net capital because its terminal exposure is nearly hedged. But
	the trade is not capital-free along the path. If the index rises, Bob's short
	futures leg generates an immediate variation-margin cash outflow, while the
	offsetting option gain is not paid into the futures margin account as same-day
	cash. If the index later falls, the futures leg generates a cash inflow that
	can be withdrawn or invested. Thus, parity enforcement is not only a terminal
	payoff identity; it is also a pre-maturity cash-management problem. A parity
	arbitrage can be cheap to enter, but costly to survive.
	
	This paper studies that survival-capital channel. The relevant burden is not
	only the initial margin assigned at inception, but the pathwise cash support
	required to keep the position alive under daily variation settlement. Interim
	cash deficits must be funded, and interim cash surpluses carry an opportunity
	cost. The cost of maintaining the trade therefore depends jointly on the scale
	of interim cash-flow exposure and the price of funding liquidity.
	
	To measure this channel, I extract option-implied discount factors from
	minute-level NBBO data on S\&P~500 and Russell~2000 options, following the
	identification approach of \citet{AB21}. I then construct the carry gap by
	comparing these option-implied discount factors with a bootstrapped OIS curve.
	
	The empirical results have three main features. First, the carry gap is
	centered in positive territory in both markets, with distributional properties
	difficult to attribute to microstructure noise or measurement error alone.
	Second, it retains pronounced low-frequency structure and regime-dependent
	persistence after daily aggregation. Third, these patterns survive
	out-of-sample validation, residual-stationarity diagnostics for the baseline
	level relation, and an alternative-benchmark analysis using Treasury
	constant-maturity yields.
	
	To interpret these facts, I introduce a reduced-form path-risk term motivated
	by geometric Brownian motion. If parity-enforcement positions require interim
	cash support before maturity, the survival-capital requirement should increase
	with volatility and time to maturity. The Brownian support-capital argument
	implies a volatility \(\times\sqrt{\tau}\) scaling. Multiplying this scale by
	a rate-like opportunity-cost variable converts survival cash into a carry
	cost. The resulting term is not a price of the terminal payoff, but a
	reduced-form proxy for the cost of surviving the pre-maturity
	variation-margin path.
	
	I place this path-risk term at the center of the regression specification,
	supplemented by trading-friction and broad financial-condition variables. The
	specification has meaningful in-sample explanatory power, and the signs of all
	key coefficients remain stable across leave-one-year-out (LOYO) out-of-sample
	validation. I interpret this evidence as reduced-form support for a pathwise
	funding and survival-capital channel, not as a structural margin model.
	
	The paper makes three contributions. First, it separates the terminal-payoff
	identity of put--call parity from its implementation as a trading strategy,
	showing that a small quoted parity residual does not automatically imply
	risk-free enforcement. Second, it documents that the carry gap---the annualized
	difference between option-implied and OIS-benchmark discount factors---is a
	systematic empirical object with a positive center, low-frequency persistence,
	and state dependence. Third, it shows that a Brownian support-capital proxy,
	together with trading-friction and financial-condition variables, forms a
	central explanatory block for this carry-space wedge. The apparent return to
	parity enforcement is therefore better understood not as a literal free lunch,
	but as compensation consistent with implementation risk, cash-flow timing, and
	finite capital.
	
	The remainder of the paper is organized as follows.
	Section~\ref{sec:lit} reviews the related literature.
	Section~\ref{sec:data} describes the data and methodology.
	Section~\ref{sec:result} presents the carry-gap estimates.
	Section~\ref{sec:gbm_reg} introduces the reduced-form regression specification.
	Sections~\ref{sec:is} and~\ref{sec:oos} report in-sample results and LOYO
	out-of-sample validation.
	Section~\ref{sec:robust} presents robustness checks, including
	residual-stationarity diagnostics and an alternative-benchmark analysis using
	Treasury constant-maturity yields in place of OIS.
	Section~\ref{sec:discuss} discusses economic implications and limitations, and
	Section~\ref{sec:conclusion} concludes.
	Appendix~\ref{app:alt} constructs an alternative spot--futures carry-gap route
	that avoids the option cross-section and compares its GBM structure with the
	AB21 option-implied route.
	Additional maturity-bin-level time-series fit results are provided in
	Appendix~\ref{app:additional_fig}.
	
	\section{Related Literature}
	\label{sec:lit}
	
	This paper connects to three strands of literature: put--call parity,
	limits to arbitrage, and option-implied discounting. The common issue is that
	put--call parity is exact as a terminal-payoff relation, but its empirical
	implementation depends on trading conditions, funding conditions, and capital
	constraints. I build most directly on the implied-discount-factor literature,
	but reinterpret the resulting deviation as a dynamic carry-space object rather
	than as a static pricing discrepancy.
	
	The classic parity literature shows that observed deviations from put--call
	parity should not be read mechanically as failures of no-arbitrage logic.
	\citet{Stoll69} noted that transaction costs, short-sale constraints, and
	dividend uncertainty can turn the frictionless equality into a no-arbitrage
	band. Subsequent empirical work examined whether apparent deviations represent
	genuine arbitrage opportunities or residuals generated by execution costs and
	market frictions \citep{GG74,KR79,AT01}. This paper shares that concern, but
	shifts the object of analysis. The question is not whether visible price-space
	parity deviations exist, but whether a systematic carry-space wedge remains
	after those deviations have largely been compressed.
	
	This distinction connects the paper to the limits-to-arbitrage literature.
	Following \citet{SV97}, this literature emphasizes that arbitrage is performed
	by capital-constrained intermediaries and is exposed to funding constraints,
	margin requirements, interim losses, and path-dependent payoffs
	\citep{GV02,BP09,MP12}. In options markets, \citet{ORW04} show that put--call
	parity violations are larger and more frequent when short-sale constraints are
	more severe. My setting is deliberately different. I study European index
	options, where quoted parity residuals against traded futures are already
	tightly compressed and the futures leg provides a direct enforcement
	instrument. The object is therefore not an unclosed price-space arbitrage
	violation, but a shadow cost of successful parity enforcement that remains
	visible in carry space.
	
	A separate strand uses option prices to infer implied interest rates or funding
	conditions. \citet{BG86} infer implied interest rates from option prices, and
	\citet{AB21} estimate option-implied discount factors from European put--call
	parity and compare them with the OIS curve. I inherit the identification logic
	of \citet{AB21}, but extend the analysis to a longer time series and a richer
	SPX--RUT panel. More importantly, I place the time-series structure of the
	deviation at the center of the analysis and ask whether it is systematically
	related to implementation risk, trading frictions, and financial conditions.
	
	The contribution is therefore not to document another unclosed arbitrage
	violation. It is to show that, even in a clean index-option setting where
	visible put--call parity residuals are nearly eliminated, enforcing the parity
	relation can leave a systematic, state-dependent wedge in carry space.
	
	\section{Data and Methodology}
	\label{sec:data}
	
	\subsection{Data and sample scope}
	
	I extract market-implied discount factors from SPX and RUT options
	and compare them with OIS discount factors to measure the carry gap.
	The identification follows the synthetic-forward procedure of \citet{AB21},
	which recovers the market-implied discount factor from European call and put
	prices at the same maturity alone.
	
	This within-option identification is central to the empirical design.
	A direct spot--futures carry construction requires combining spot, futures,
	dividend, and interest-rate inputs, and is therefore exposed to
	nonsynchronicity, dividend-estimation error, and benchmark-matching noise.
	By contrast, the \citet{AB21} procedure identifies the discount factor inside
	the option cross-section itself: the forward and discount factor are estimated
	jointly from call--put spreads across strikes. This makes the main carry-gap
	measure substantially cleaner than the spot--futures route examined in
	Appendix~\ref{app:alt}, which I use as an external robustness diagnostic rather
	than as the baseline measure.
	
	Option quotes are minute-level NBBO data collected from ThetaData.
	Although option data are available through December~31, 2025, the analysis
	sample is restricted to January~4, 2016 through October~31, 2025 to match the
	availability of OIS data. All results are based on the common sample in which
	option-market information and OIS discount curves are simultaneously
	observable.
	
	Both SPX and RUT options are European-style index options, so early-exercise
	premia do not introduce institutional noise into parity-based
	discount-factor identification.
	
	All empirical analysis is conducted in MATLAB R2025b.\footnote{%
		On 16 parallel workers,
		the full pipeline for both markets executes in approximately one hour.}
	
	\subsection{Identification of option-implied discount factors}
	
	The identification logic follows \citet{AB21}.
	For a European call and put at strike $K$ with maturity $T$
	observed at time $t$, put--call parity can be written as
	\begin{equation}
		C_t(K,T) - P_t(K,T) = B_t(T)\bigl(F_t(T) - K\bigr),
	\end{equation}
	where $B_t(T)$ is the market-implied discount factor
	and $F_t(T)$ is the forward value at the same maturity.
	
	Defining the synthetic forward as
	\begin{equation}
		\mathcal{G}_t(K,T) = C_t(K,T) - P_t(K,T),
	\end{equation}
	no-arbitrage requires the recovered forward value to be independent of $K$.
	The market-implied discount factor is therefore the value that makes
	\begin{equation}
		F_t(T) = \frac{\mathcal{G}_t(K,T)}{B_t(T)} + K
	\end{equation}
	constant across strikes.
	
	In practice, for each date--maturity pair I exploit the linear relation
	between the synthetic forward and the strike to estimate $\hat{B}_t(T)$ and
	$\hat{F}_t(T)$ simultaneously. The option-implied discount factor is the
	discount rate that eliminates strike-dependence in the forward price recovered
	from the synthetic forward.
	
	This approach has three advantages. First, it uses the full strike
	cross-section within a maturity, reducing dependence on a specific ATM contract
	or arbitrary moneyness range. Second, because $\hat{B}_t(T)$ and $\hat{F}_t(T)$
	are identified jointly, dividends are absorbed into the recovered forward and
	need not be estimated separately. Third, all identifying prices come from the
	same option cross-section, mitigating the nonsynchronicity problems that arise
	when spot, futures, dividend, and rate data are combined externally.
	
	I apply this procedure across the SPX and RUT samples to construct a
	market$\times$date$\times$maturity panel of implied discount factors.
	
	\subsection{OIS curve construction and carry-gap definition}
	
	The benchmark discount factor is derived from the OIS curve. Since the
	financial crisis, OIS has become the standard benchmark for derivatives
	discounting, and \citet{AB21} likewise measure funding spreads against it.
	
	I bootstrap daily OIS data to recover maturity-matched discount factors and
	zero rates, and construct maturity-matched OIS discount factors for direct
	comparison with $\hat{B}_t(T)$.
	
	The carry gap is defined as the annualized deviation between the two discount
	factors. Letting $\tau_t(T)=T-t$,
	\begin{equation}
		CG_t(T)
		=
		\frac{1}{\tau_t(T)}
		\log\!\left(
		\frac{D_t^{\mathrm{OIS}}(T)}{\hat{B}_t(T)}
		\right),
		\label{eq:carry_gap_def}
	\end{equation}
	where $D_t^{\mathrm{OIS}}(T)$ is the OIS discount factor
	and $\hat{B}_t(T)$ is the option-implied discount factor.
	$CG_t(T)>0$ indicates that the options market embeds a higher implied carry
	than the OIS benchmark.
	
	The empirical analysis uses the basis-point-scaled version
	\begin{equation}
		CG_t^{bp}(T) = 10^4 \cdot CG_t(T),
		\label{eq:carry_gap_bp_def}
	\end{equation}
	and the daily, market-level carry gap entering regressions is denoted
	$CG_{i,t}^{bp}$.
	
	\subsection{Sample filters and final panel construction}
	
	The preprocessing removes observations with low liquidity or unstable price
	information to ensure stable cross-sectional identification. Only call--put
	pairs sharing the same strike and maturity are used. I exclude observations
	with abnormally low prices or excessive bid--ask spreads, maturities with too
	few valid strikes, and dates on which OIS curve recovery fails or the term
	structure is anomalous. The final sample consists of observations for which
	(i)~the option-implied discount factor can be identified and
	(ii)~the OIS discount factor can be reliably constructed at the same date and
	maturity.
	
	I construct a date$\times$maturity panel for each of SPX and RUT.
	Daily time series are aggregated as the median of eligible observations on
	each date, reducing sensitivity to outliers and transient noise while tracking
	the central movement of the carry gap.

	\section{Carry-Gap Estimates}
	\label{sec:result}
	
	\FloatBarrier
	\begin{figure}[H]
		\centering
		\includegraphics[width=6.5in]{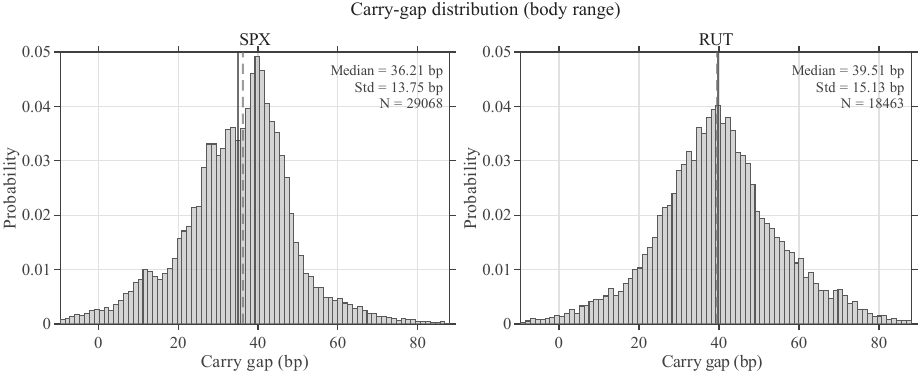}
		\caption{Distribution of daily carry gaps for SPX and RUT.
			Both distributions are centered in positive territory rather than at zero.
			RUT is roughly bell-shaped, while SPX displays stronger asymmetry.}
		\label{fig:hist}
		
		\vspace{0.25em}
		\begin{minipage}{0.90\linewidth}
			\footnotesize
			\textit{Notes:} The histograms are plotted over the body range shown
			on the horizontal axis. The $N$ inside each panel denotes observations
			within this plotted range, not the full date--maturity regression sample.
		\end{minipage}
	\end{figure}
	\FloatBarrier
	
	\subsection{Center and cross-market shape}
	
	Figure~\ref{fig:hist} shows that the carry-gap distribution is shifted into
	positive territory in both markets. The full-sample mean is 36.91\,bp, the
	median is 37.50\,bp, and 98.4\% of observations are positive. By market, the
	mean and median are 34.87\,bp and 36.16\,bp for SPX, and 40.12\,bp and
	39.57\,bp for RUT. Thus, the positive carry gap is not confined to one market.
	
	The shapes differ across markets. RUT is relatively smooth and bell-shaped,
	whereas SPX has a sharper peak and a longer right tail. This heterogeneity is
	consistent with differences in liquidity and implementation environment, while
	the positive center is common to both.
	
	I also verify that the positive center is not an artifact of poor
	cross-sectional fitting. The date--maturity regressions in the \citet{AB21}
	pipeline achieve near-perfect fit: the median cell-level $R^2$ is 0.9999999
	for SPX and 0.9999995 for RUT, with minimum values of 0.9999972 and
	0.9999848, respectively.\footnote{%
		For some date--maturity combinations, too few observations survive
		preprocessing for the regression to be estimated. These cases reflect
		insufficient information for identification rather than poor-quality fits.}
	If the carry gap were zero-centered noise, daily aggregation should move the
	distribution toward zero. Instead, negative observations are rare and the
	distribution remains shifted into positive territory.
	
	\subsection{Magnitude and maturity structure}
	\label{subsec:result_size_maturity}
	
	The full-sample median of approximately 37\,bp is close to the roughly 34\,bp
	reported by \citet{AB21}. The samples and measurement details are not identical,
	so the comparison should not be mechanical, but the proximity is a useful sanity
	check.
	
	\FloatBarrier
	\begin{figure}[H]
		\centering
		\includegraphics[width=6.5in]{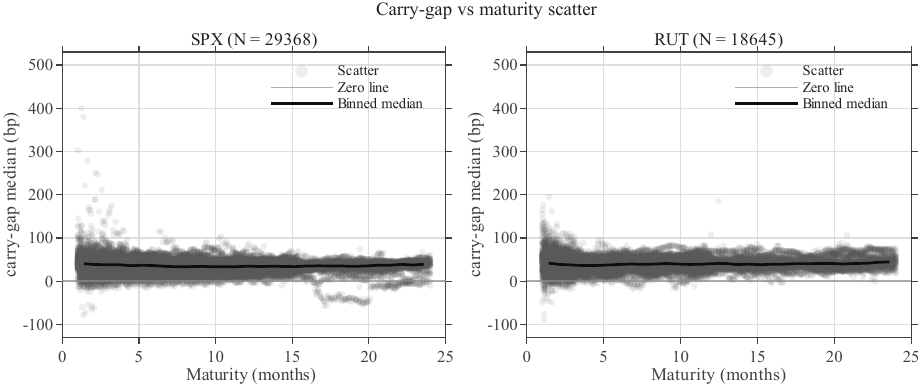}
		\caption{Carry gaps against time to maturity.
			Each point is a date--maturity observation; the bold line is the binned
			median. Binned medians remain positive across maturities, while dispersion
			is much wider at the short end.}
		\label{fig:vs_tau}
	\end{figure}
	\FloatBarrier
	
	Figure~\ref{fig:vs_tau} shows two maturity patterns. First, binned medians
	remain positive across the maturity spectrum, roughly in the 30--40\,bp range.
	The carry gap is therefore not a short-maturity-only distortion. Second,
	dispersion is strongly maturity-dependent: short-maturity observations are much
	more dispersed, with outliers above 400\,bp, but the scatter compresses rapidly
	as maturity increases. This reflects both annualization through a small
	$\tau$ and the greater influence of microstructure frictions at short horizons.
	The combination of a flat level profile and maturity-dependent dispersion
	motivates the maturity-sensitive specification in Section~\ref{sec:gbm_reg}.
	
	\subsection{Time-series variation}
	\label{subsec:result_ts}
	
	\FloatBarrier
	\begin{figure}[H]
		\centering
		\includegraphics[width=6.5in]{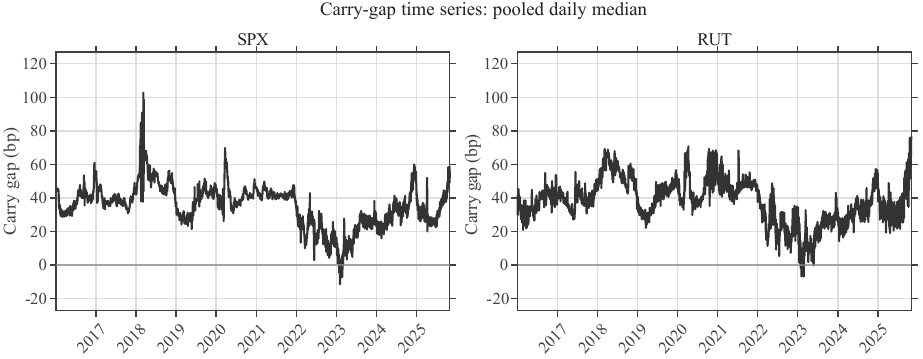}
		\caption{Daily carry-gap time series for SPX and RUT.
			Each value is the pooled daily median across eligible observations.
			The carry gap is positive for most of the sample and exhibits persistent
			level shifts.}
		\label{fig:cg_series}
	\end{figure}
	\FloatBarrier
	
	Figure~\ref{fig:cg_series} shows that the carry gap is also a time-series
	object. It remains positive for most of the sample and exhibits persistent
	level shifts rather than quick reversion to zero. Both markets display elevated
	levels in 2018 and 2020--2021, a decline during 2022--2023, and a rebound in
	2024--2025. The detailed paths differ, but the broad regime shifts are shared,
	suggesting a systematic component linked to the broader market environment.
	
	In sum, the carry gap has three main properties: a positive center, a relatively
	flat maturity profile, and low-frequency time-series variation. These facts
	motivate treating it as a systematic carry-space object rather than residual
	noise.
	
	\section{Path-Risk Term and Regression Specification}
	\label{sec:gbm_reg}
	
	This section introduces the empirical specification for the carry gap.
	The key regressor is a GBM-based path-risk term that captures the
	pre-maturity implementation burden of parity enforcement rather than the
	terminal payoff itself. Put--call parity is exact at maturity, but the
	enforcing strategy must survive daily settlement, variation margin, interim
	losses, and finite-capital constraints. The enforcement problem is therefore
	path-dependent even when the terminal payoff is deterministic.
	
	This distinction reconciles two facts: quoted parity residuals are tightly
	compressed in price space, yet a systematic wedge persists in carry space.
	If arbitrage compresses visible residuals while leaving implementation risk in
	the enforcement process, the remaining wedge should be related to variables
	governing pathwise funding and capital support. To my knowledge, prior work
	has not placed an explicit functional form for parity-enforcement path risk at
	the center of a carry-gap regression.
	
	\subsection{Intuition and derivation of the path-risk term}
	
	The GBM term begins from the observation that a parity-enforcement position can
	be deterministic at maturity but capital-using before maturity. A position
	combining a synthetic long forward with a short futures contract pays
	\[
	(S_T - K) + (F_0 - S_T) = F_0 - K
	\]
	at maturity. Before maturity, however, futures-price movements affect the
	margin account. Adverse paths require capital support, and failure to provide
	that support can prevent the trader from reaching the terminal payoff.
	
	Let the normalized interim P\&L process be
	\[
	X_t = \sigma B_t,
	\]
	where $B_t$ is standard Brownian motion and $\sigma$ is annualized volatility.
	Under the minimal support rule that keeps the position solvent, cumulative
	support capital $L_t$ satisfies
	\[
	X_t + L_t \ge 0
	\qquad \text{for all } t \in [0,T],
	\qquad L_0 = 0,
	\]
	with the smallest such nondecreasing process given by
	\[
	L_t = \sup_{0 \le s \le t}(-X_s)^+.
	\]
	
	By standard Brownian-motion properties,
	\[
	\mathbb{E}[L_t / N]
	=
	\sigma\sqrt{\frac{2t}{\pi}},
	\]
	so expected support capital is proportional to volatility and rises with the
	square root of time. Averaging over the life of the trade gives
	\[
	\bar{B}(T)
	=
	\frac{1}{T}\int_0^T \mathbb{E}[L_t / N]\,dt
	=
	\frac{2}{3}\,\sigma\sqrt{\frac{2T}{\pi}}.
	\]
	
	If the opportunity cost of committed capital is summarized by a rate-like
	object $r_t$, the representative path-risk scale is
	\[
	r_t\,\bar{B}(T)
	=
	r_t \cdot \frac{2}{3}\,\sigma\sqrt{\frac{2T}{\pi}}.
	\]
	The empirical GBM term translates this scale into basis points:
	\begin{equation}
		\label{eq:gbm_term_definition}
		\begin{aligned}
			GBM_{i,t}^{OIS,xY}
			&=
			10^4 \cdot
			\frac{OISxY_t}{100}
			\cdot
			\frac{2}{3}
			\cdot
			\frac{Vol_{i,t}}{100}
			\cdot
			\sqrt{\frac{2\tau_{i,t}}{\pi}},
			\qquad x \in \{1,10\},
			\\[0.5em]
			Vol_{i,t}
			&=
			\begin{cases}
				VIX_t, & i = \mathrm{SPX},\\
				RVX_t, & i = \mathrm{RUT}.
			\end{cases}
		\end{aligned}
	\end{equation}
	The $x=1$ component proxies for short-to-medium-term funding conditions, while
	$x=10$ proxies for the long-run opportunity cost of capital. The term is not a
	structural margin model; it is a reduced-form proxy for path-dependent capital
	support.
	
	\subsection{Regression specifications}
	
	The baseline specification pools SPX and RUT to estimate the average GBM
	structure common to both markets, while allowing a mean-level difference through
	an SPX dummy. The estimation sample is restricted to observations with at least
	one month to maturity:
	\begin{equation}
		\label{eq:pooled_gbm_spec}
		\begin{aligned}
			CG_{i,t}^{bp}
			=
			{}&
			\alpha
			+
			\delta\, D_i^{\mathrm{SPX}}
			+
			\phi_{1}\,GBM_{i,t}^{OIS,1Y}
			+
			\phi_{10}\,GBM_{i,t}^{OIS,10Y}
			\\
			&+
			\beta\,\frac{BA^{\mathrm{med}}_{i,t}}{\tau_{i,t}}
			+
			\gamma\,NFCI_t
			+
			\varepsilon_{i,t},
		\end{aligned}
	\end{equation}
	where $D_i^{\mathrm{SPX}}=1$ for SPX and zero for RUT.
	
	The two GBM terms form the core explanatory block. The bid--ask term captures
	trading frictions and has a larger annualized effect at shorter maturities,
	while NFCI proxies for system-wide funding stress and financial conditions.
	The GBM coefficients are not fixed at one because parity can be enforced from
	both sides. The $+C-P-F$ and $-C+P+F$ trades share the same path-risk scale but
	have opposite economic directions. The observed carry gap is therefore
	interpreted as a net directional imbalance between opposing enforcement
	pressures.
	
	Because SPX and RUT differ in liquidity, investor base, and microstructure, I
	also estimate market-specific regressions:
	\begin{equation}
		\label{eq:market_specific_gbm_spec}
		CG_{i,t}^{bp}
		=
		\alpha_i
		+
		\phi_{1,i}\,GBM_{i,t}^{OIS,1Y}
		+
		\phi_{10,i}\,GBM_{i,t}^{OIS,10Y}
		+
		\beta_i\,\frac{BA^{\mathrm{med}}_{i,t}}{\tau_{i,t}}
		+
		\gamma_i\,NFCI_t
		+
		\varepsilon_{i,t},
		\qquad
		i \in \{\mathrm{SPX},\mathrm{RUT}\}.
	\end{equation}
	The pooled regression summarizes the common structure, while the market-specific
	regressions reveal heterogeneity in the loadings.
	
	In sum, the specification explains the carry gap through a GBM path-risk block,
	supplemented by trading-friction and financial-condition variables. The GBM
	terms are the central explanatory objects, not auxiliary controls.
		
	\section{In-Sample Results}
	\label{sec:is}
	
	This section examines the in-sample explanatory power of the GBM-based
	reduced-form specification. The focus is not on replicating every daily
	observation, but on overall fit, coefficient-sign stability, and whether the
	specification summarizes the central structure of the carry gap.
	
	I compare three specifications: a pooled common-market specification with
	common slopes and an SPX dummy, and separate specifications for SPX and RUT.
	Because the carry gap is measured daily and exhibits persistent low-frequency
	variation, coefficient inference uses date-based HAC (Newey--West) standard
	errors with a maximum lag of 21 trading days. For a sample of roughly 2,456
	trading days, a standard automatic lag rule would select about eight lags, so
	the fixed 21-day choice is deliberately conservative and allows for residual
	dependence over approximately one trading month.
	
	\FloatBarrier
	\begin{table}[H]
		\centering
		\footnotesize
		\caption{In-sample fit summary}
		\label{tab:is_summary_gbm}
		\resizebox{6.5in}{!}{
			\begin{tabular}{lcccccc}
				\toprule
				Specification & Obs. & Trading days & $R^2$ & Adj.\ $R^2$ & RMSE (bp) & MAE (bp) \\
				\midrule
				Pooled common + SPX dummy & 48{,}013 & 2{,}456 & 0.309 & 0.309 & 13.57 & 9.26 \\
				SPX separate              & 29{,}368 & 2{,}456 & 0.312 & 0.312 & 13.20 & 8.68 \\
				RUT separate              & 18{,}645 & 2{,}455 & 0.281 & 0.281 & 13.95 & 10.10 \\
				\bottomrule
			\end{tabular}
		}
	\end{table}
	\FloatBarrier
	
	\FloatBarrier
	\begin{table}[H]
		\centering
		\footnotesize
		\caption{Coefficient estimates with HAC(21) inference}
		\label{tab:is_coef_gbm_hac21}
		\begin{tabular}{lccc}
			\toprule
			Regressor 
			& Pooled common + SPX dummy 
			& SPX separate 
			& RUT separate \\
			\midrule
			
			Intercept
			& 24.901*** & 23.134*** & 24.577*** \\
			& (5.816)   & (5.713)   & (5.407)   \\
			
			$D^{\mathrm{SPX}}$
			& -0.985    & ---       & ---       \\
			& (0.713)   &           &           \\
			
			$GBM^{OIS,1Y}$
			& -0.557*** & -0.548*** & -0.555*** \\
			& (0.148)   & (0.170)   & (0.124)   \\
			
			$GBM^{OIS,10Y}$
			& 0.469***  & 0.411**   & 0.541***  \\
			& (0.151)   & (0.172)   & (0.130)   \\
			
			$BA^{\mathrm{med}}/\tau$
			& 0.158***  & 0.256***  & 0.130***  \\
			& (0.029)   & (0.064)   & (0.022)   \\
			
			$NFCI$
			& -24.598** & -25.839** & -23.961** \\
			& (10.283)  & (10.359)  & (10.013)  \\
			
			\midrule
			Trading days 
			& 2{,}456 & 2{,}456 & 2{,}455 \\
			\bottomrule
		\end{tabular}
		
		\vspace{0.5em}
		\begin{minipage}{0.95\linewidth}
			\footnotesize
			\textit{Notes:} Standard errors in parentheses are date-based HAC
			(Newey--West) standard errors with maximum lag 21 trading days.
			In the pooled common specification, $D^{\mathrm{SPX}}$ is an indicator equal to one for SPX
			and zero for RUT, so the intercept corresponds to the RUT level.
			***, **, * denote significance at the 1\%, 5\%, and 10\% levels, respectively.
		\end{minipage}
	\end{table}
	\FloatBarrier
	
	\subsection{Overall fit and coefficient structure}
	
	Table~\ref{tab:is_summary_gbm} shows that the reduced-form specification has
	meaningful explanatory power. The pooled common-slope specification has an
	$R^2$ of 0.309 and an RMSE of 13.57\,bp. The separate specifications deliver
	similar fit, with $R^2$ values of 0.312 for SPX and 0.281 for RUT. The limited
	gain from estimating separate models suggests that the central structure of the
	carry gap is largely common across the two markets.
	
	Table~\ref{tab:is_coef_gbm_hac21} shows that the coefficient pattern is stable
	across specifications. The short-horizon GBM term is negative, the long-horizon
	GBM term is positive, the bid--ask term is positive, and NFCI is negative. All
	four variables are statistically significant in the pooled specification under
	HAC(21) inference, and the same sign pattern is preserved in the market-specific
	regressions. The SPX dummy is small and insignificant, indicating that
	mean-level differences between SPX and RUT are not the main source of
	explanatory power once path-risk, trading-friction, and financial-condition
	variables are included.
	
	The opposite signs of $GBM^{OIS,1Y}$ and $GBM^{OIS,10Y}$ are consistent with
	the interpretation that the carry gap reflects a net directional imbalance
	between opposing parity-enforcement pressures, rather than the total path cost
	of one-sided enforcement. The GBM coefficients are therefore reduced-form
	loadings of net enforcement pressure on distinct rate-like components of
	path-risk.
	
	\subsection{Market heterogeneity and maturity-bin fit}
	
	The pooled estimates are close to the market-specific estimates for the
	short-horizon GBM term and NFCI, reinforcing the view that these variables
	capture a common structure rather than a market-specific artifact. Heterogeneity
	appears mainly in the long-horizon GBM and bid--ask channels. The long-horizon
	GBM term loads more strongly in RUT, whereas the bid--ask term loads more
	strongly in SPX. Thus, the separate regressions do not overturn the common
	structure; they show where market-specific sensitivities are concentrated.
	
	Explanatory power is strongest at intermediate maturities. For SPX under the
	pooled specification, $R^2$ rises from 0.080 at 1--2 months to 0.530 at
	10--14 months, then declines to 0.254 beyond 21 months. RUT shows a similar
	pattern, with $R^2$ reaching 0.440 at 10--14 months and 0.451 at 14--21 months,
	remaining at 0.363 beyond 21 months. The maturity-bin results therefore
	reinforce the main message: the specification captures a common path-risk
	structure most clearly at intermediate maturities, rather than delivering fully
	customized market-level fits.
	
	\subsection{Error diagnostics and interpretation}
	
	The daily mean relative error under the pooled specification is approximately
	$-13\%$, with a mean absolute relative error of approximately $29\%$. The
	separate specifications are similar. Fitted values therefore tend to lie
	slightly below actuals, consistent with a reduced-form model of central levels
	and major regime shifts rather than a model designed to replicate every
	high-frequency fluctuation.
	
	In sum, the in-sample results support the interpretation that the GBM path-risk
	block captures the central structure of the carry gap. Under HAC(21) inference,
	$GBM^{OIS,1Y}$ remains negative, $GBM^{OIS,10Y}$ remains positive,
	$BA^{\mathrm{med}}/\tau$ remains positive, and NFCI remains negative in both
	pooled and market-specific specifications. The pooled specification is not
	materially inferior to the separate specifications, while the market-specific
	estimates reveal economically meaningful heterogeneity in the long-run GBM and
	bid--ask channels.
	
	\section{Out-of-Sample Validation}
	\label{sec:oos}
	
	This section evaluates whether the GBM-based structure is tied to particular
	calendar years. I use a leave-one-year-out (LOYO) procedure: each calendar year
	is held out in turn, coefficients are estimated on all remaining years, and fit
	is evaluated on the excluded year. The advantage of this design is that every
	year in the sample can serve as a validation period, including crisis,
	transition, and stabilization regimes.
	
	The LOYO exercise should not be interpreted as a causal, real-time forecasting
	test. Because the training sample for a given holdout year includes observations
	both before and after that year, the procedure uses information that would not
	have been available to an investor forecasting forward in calendar time. A
	strictly time-ordered expanding-window design would be closer to such a
	real-time forecasting exercise. In the present sample, however, the available
	post-GFC OIS-overlap window covers only about nine years and ten months. Once a
	reasonable initial training window is reserved, only a small number of calendar
	years remains for evaluation; using a very short initial window would instead
	make coefficient estimates unstable. I therefore use LOYO for the narrower
	purpose of testing whether the reduced-form relation is an artifact of
	overfitting to particular years, or whether the same structure remains useful
	when any one year is excluded from estimation.
	
	The evaluation focuses on two questions. First, does out-of-sample fit fail
	broadly, or is weakness concentrated in particular regimes? Second, do the
	re-estimated coefficients preserve their signs and significance across LOYO
	training samples? Coefficient-stability diagnostics are reported for the
	market-specific baseline regressions, using date-based HAC (Newey--West)
	standard errors with a maximum lag of 21 trading days.\footnote{%
		For samples of this size, the common automatic rule
		$\lfloor 4(T/100)^{2/9}\rfloor$ selects roughly eight lags.
		The fixed 21-day choice is therefore deliberately conservative and
		allows for residual dependence over approximately one trading month.}
	
	\subsection{LOYO design and evaluation criteria}
	
	The sample spans the pandemic shock, a rapid rate-hiking cycle, and the
	subsequent stabilization. LOYO validation is useful in this setting because it
	allows every calendar year to serve as a holdout regime without committing the
	analysis to a single arbitrary split. This is especially useful in a short
	post-GFC sample: a strictly expanding-window exercise would be more causal in
	calendar time, but would leave too few independent annual evaluation periods
	after reserving a credible initial training window.
	
	I use year-level out-of-sample $R^2$ as the primary metric, interpreted
	together with mean $R^2$, median $R^2$, pooled $R^2$, the number of years with
	positive $R^2$, correlation, and RMSE. Coefficient stability is evaluated
	separately using the signs and HAC(21) significance of coefficients
	re-estimated within each LOYO training sample.
	
	\subsection{Results}
	
	\FloatBarrier
	\begin{table}[H]
		\centering
		\footnotesize
		\caption{LOYO out-of-sample performance summary}
		\label{tab:oos_summary_gbm}
		\resizebox{6.5in}{!}{
			\begin{tabular}{llcccccc}
				\toprule
				Specification & Market & Mean $R^2$ & Median $R^2$ & Pooled $R^2$ & Years with $R^2>0$ & Mean corr. & Mean RMSE (bp) \\
				\midrule
				Common-market & SPX & 0.049 & 0.187 & 0.212 & 9/10 & 0.189 & 13.93 \\
				Common-market & RUT & 0.065 & 0.063 & 0.173 & 6/10 & 0.252 & 15.16 \\
				Separate      & SPX & 0.059 & 0.130 & 0.221 & 7/10 & 0.205 & 13.95 \\
				Separate      & RUT & 0.075 & 0.108 & 0.171 & 6/10 & 0.243 & 15.07 \\
				\bottomrule
			\end{tabular}%
		}
	\end{table}
	\FloatBarrier
	
	Table~\ref{tab:oos_summary_gbm} shows modest mean out-of-sample $R^2$
	values. Mean $R^2$ ranges from 0.049 to 0.075 across the four
	specification--market combinations, so the specification should not be read
	as a strong forecasting model. At the same time, performance does not fail
	uniformly: positive $R^2$ is recorded in 9 of 10 years for SPX and 6 of 10
	years for RUT under the common-market specification, and pooled $R^2$ remains
	positive in both markets.
	
	The weakness is concentrated in a small number of holdout years. For SPX, the
	2020 holdout is the dominant failure case, with $R^2=-1.634$ under the
	common-market specification and $R^2=-1.221$ under the separate
	specification. For RUT, 2020 and the 2016--2017 holdouts also generate
	negative $R^2$. By contrast, several post-2020 holdouts perform well; for
	example, the 2023 common-market $R^2$ is 0.561 for SPX and 0.664 for RUT.
	
	\FloatBarrier
	\begin{table}[H]
		\centering
		\footnotesize
		\caption{LOYO out-of-sample performance excluding the 2020 holdout}
		\label{tab:oos_ex2020_gbm}
		\resizebox{6.5in}{!}{
			\begin{tabular}{llccccc}
				\toprule
				Specification & Market & Mean $R^2$ & Median $R^2$ & Years with $R^2>0$ & Mean corr. & Mean RMSE (bp) \\
				\midrule
				Common-market & SPX & 0.236 & 0.215 & 9/9 & 0.250 & 13.25 \\
				Common-market & RUT & 0.119 & 0.091 & 6/9 & 0.263 & 14.49 \\
				Separate      & SPX & 0.201 & 0.185 & 7/9 & 0.261 & 13.45 \\
				Separate      & RUT & 0.148 & 0.153 & 6/9 & 0.258 & 14.26 \\
				\bottomrule
			\end{tabular}
		}
	\end{table}
	\FloatBarrier
	
	Table~\ref{tab:oos_ex2020_gbm} confirms that the average weakness is driven
	partly by the extreme 2020 regime. Excluding the 2020 holdout, mean $R^2$
	rises to 0.236 for SPX under the common-market specification and 0.201 under
	the separate specification. For RUT, the corresponding values rise to 0.119
	and 0.148. Mean correlations also stabilize in the 0.24--0.26 range, and RMSE
	stabilizes around 13.3--13.5\,bp for SPX and 14.3--14.5\,bp for RUT. The main
	instability is therefore level calibration in specific regimes, not a complete
	loss of directionality.
	
	\FloatBarrier
	\begin{table}[H]
		\centering
		\footnotesize
		\caption{LOYO coefficient stability with HAC(21) inference}
		\label{tab:oos_sign_gbm_hac21}
		\resizebox{6.5in}{!}{
			\begin{tabular}{lcccc}
				\toprule
				Regressor
				& SPX sign
				& SPX HAC significance
				& RUT sign
				& RUT HAC significance \\
				\midrule
				Intercept
				& $+$\;10/10 & 10/10 at 1\%
				& $+$\;10/10 & 10/10 at 1\% \\
				
				$GBM^{OIS,1Y}$
				& $-$\;10/10 & 9/10 at 5\%
				& $-$\;10/10 & 10/10 at 1\% \\
				
				$GBM^{OIS,10Y}$
				& $+$\;10/10 & 8/10 at 5\%
				& $+$\;10/10 & 10/10 at 1\% \\
				
				$BA^{\mathrm{med}}/\tau$
				& $+$\;10/10 & 10/10 at 1\%
				& $+$\;10/10 & 10/10 at 1\% \\
				
				$NFCI$
				& $-$\;10/10 & 8/10 at 5\%, 10/10 at 10\%
				& $-$\;10/10 & 9/10 at 5\%, 10/10 at 10\% \\
				\bottomrule
			\end{tabular}
		}
		
		\vspace{0.5em}
		\begin{minipage}{0.92\linewidth}
			\footnotesize
			\textit{Notes:} Each entry is computed from the ten leave-one-year-out
			training-sample regressions for the market-specific baseline specification.
			Standard errors are date-based HAC (Newey--West) standard errors
			with maximum lag 21 trading days.
		\end{minipage}
	\end{table}
	\FloatBarrier
	
	Table~\ref{tab:oos_sign_gbm_hac21} shows that the coefficient structure does
	not collapse across LOYO folds. In both markets, the signs of all four
	non-intercept key regressors are stable in 10 out of 10 re-estimated training
	samples: $GBM^{OIS,1Y}$ remains negative, $GBM^{OIS,10Y}$ remains positive,
	$BA^{\mathrm{med}}/\tau$ remains positive, and NFCI remains negative.
	
	The HAC(21) significance pattern is also supportive. For RUT, the two GBM
	terms and the bid--ask term are significant at the 1\% level in all ten
	folds, while NFCI is significant at the 5\% level in 9 out of 10 folds and at
	the 10\% level in all folds. For SPX, $GBM^{OIS,1Y}$ is significant at the
	5\% level in 9 out of 10 folds, $GBM^{OIS,10Y}$ in 8 out of 10 folds, and the
	bid--ask term in all folds. NFCI remains negative in all folds and is
	significant at the 5\% level in 8 out of 10 folds and at the 10\% level in all
	folds.
	
	Thus, weak LOYO fit in some years is not driven by random sign reversal or
	coefficient collapse. The coefficient directions remain stable even under the
	conservative HAC(21) inference, while the main instability lies in coefficient
	magnitudes and level calibration during specific regimes.
	
	\FloatBarrier
	\begin{figure}[H]
		\centering
		\includegraphics[width=6.5in]{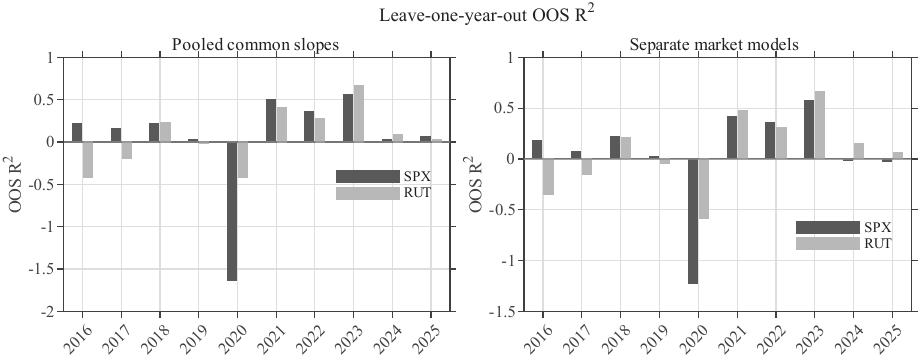}
		\caption{Year-level LOYO out-of-sample $R^2$
			for the common-market and separate specifications.
			Most holdout years produce positive or near-zero $R^2$,
			but the 2020 holdout for SPX
			and a few early holdout years for RUT
			generate sharply negative values
			that drag down the overall mean.}
		\label{fig:oos}
	\end{figure}
	\FloatBarrier
	
	Figure~\ref{fig:oos} visualizes the same pattern. Most holdout years produce
	positive or near-zero $R^2$, while a few regime-specific failures dominate
	the mean. The common-market specification provides the more restrictive
	benchmark, imposing a shared coefficient structure across SPX and RUT. The
	market-specific specifications offer only modest average fit improvements,
	but allow each market to express different coefficient sensitivities.
	
	In sum, the LOYO results should be read as year-exclusion diagnostics, not as
	real-time forecasting evidence. The specification is not a strong
	high-frequency forecasting model. Rather, it is a reduced-form structure whose
	main coefficient signs survive the exclusion of every individual calendar year.
	The concentration of fit deterioration in a few holdout years, together with
	full sign stability and broad HAC(21) significance of the re-estimated
	coefficients, suggests that the specification captures a regime-dependent
	economic structure rather than a relation overfit to a particular year.
	
	\section{Robustness}
	\label{sec:robust}
	
	This section reports two robustness exercises. First, I use
	residual-stationarity diagnostics to assess whether the OIS-based level
	relation is merely a spurious alignment of persistent variables. Second, I
	replace the OIS benchmark with Treasury constant-maturity yields (DGS) to test
	whether the GBM sign structure depends mechanically on the benchmark curve.
	
	\subsection{Residual-Stationarity Diagnostics}
	\label{subsec:residual_stationarity}
	
	A natural concern in a level regression with persistent financial variables is
	that the fitted relation may reflect unrelated low-frequency trends rather than
	an economically meaningful carry-gap structure. I therefore test whether the
	baseline OIS-based GBM specification leaves a stationary equilibrium error.
	
	For each market, I aggregate the observed carry gap to the daily median series
	and estimate the long-run relation implied by the market-specific OIS-based
	specification:
	\begin{equation}
		CG^{bp}_{i,t}
		=
		a_i
		+
		b_{1,i}GBM^{OIS,1Y}_{i,t}
		+
		b_{10,i}GBM^{OIS,10Y}_{i,t}
		+
		c_i\frac{BA^{med}_{i,t}}{\tau_{i,t}}
		+
		d_iNFCI_t
		+
		u_{i,t},
		\label{eq:eg_baseline_longrun_relation}
	\end{equation}
	where the right-hand-side variables are the same state variables used in
	equation~\eqref{eq:market_specific_gbm_spec}. The object of interest is not
	forecasting performance, but whether the residual $u_{i,t}$ retains a
	nonstationary component.
	
	Table~\ref{tab:eg_residual_stationarity} reports Engle--Granger
	residual-based tests for this relation. Because the cointegrating regression
	contains four regressors, the statistics are compared with
	regressor-count-specific Engle--Granger critical values rather than standard
	ADF critical values. I report both constant-only and constant-plus-trend
	specifications.
	
	\FloatBarrier
	\begin{table}[H]
		\centering
		\footnotesize
		\caption{Residual-stationarity diagnostics for the OIS-based baseline specification}
		\label{tab:eg_residual_stationarity}
		\resizebox{6.5in}{!}{%
			\begin{tabular}{llccccccccc}
				\toprule
				Market
				& Deterministic
				& $N$
				& $k$
				& EG model
				& EG $\tau$ stat.
				& $p$-value
				& 10\% CV
				& 5\% CV
				& 1\% CV
				& Reject 1\% \\
				\midrule
				SPX
				& Constant
				& 2{,}456
				& 4
				& H1
				& -12.009
				& 0.001
				& -4.141
				& -4.427
				& -4.976
				& Yes \\
				
				SPX
				& Constant + trend
				& 2{,}456
				& 4
				& H*
				& -12.058
				& 0.001
				& -4.446
				& -4.730
				& -5.275
				& Yes \\
				
				RUT
				& Constant
				& 2{,}455
				& 4
				& H1
				& -14.101
				& 0.001
				& -4.141
				& -4.427
				& -4.976
				& Yes \\
				
				RUT
				& Constant + trend
				& 2{,}455
				& 4
				& H*
				& -14.183
				& 0.001
				& -4.446
				& -4.730
				& -5.275
				& Yes \\
				\bottomrule
			\end{tabular}%
		}
		
		\vspace{0.5em}
		\begin{minipage}{0.94\linewidth}
			\footnotesize
			\textit{Notes:} This table reports Engle--Granger residual-stationarity diagnostics
			for the daily median carry gap and the four regressors in the OIS-based baseline
			specification: $GBM^{OIS,1Y}$, $GBM^{OIS,10Y}$, $BA^{med}/\tau$, and NFCI.
			The column $k$ denotes the number of cointegrating regressors. EG statistics are
			residual-based ADF-type $\tau$ statistics and should be compared with
			Engle--Granger critical values adjusted for the number of regressors, rather than
			standard ADF critical values. The sample runs from January 4, 2016 to
			October 31, 2025.
		\end{minipage}
	\end{table}
	\FloatBarrier
	
	The rejection is strong in both markets and under both deterministic
	specifications. The EG statistics are far below the corresponding 1\% critical
	values for both SPX and RUT, so the conclusion is not marginal and does not
	depend on whether a deterministic trend is included.
	
	These diagnostics address a distinct concern from the in-sample and LOYO
	exercises. HAC inference addresses serial correlation in coefficient
	estimation; LOYO validation addresses year-level stability; the
	Engle--Granger test asks whether the baseline level relation is merely a
	spurious alignment of unrelated persistent variables. The stationary residual
	evidence suggests that it is not. I interpret this result as a
	residual-stationarity diagnostic, not as a structural claim that one
	time-invariant cointegrating vector governs all regimes.
	
	\subsection{DGS as an Alternative Benchmark}
	
	This section replaces the OIS discount benchmark with Treasury
	constant-maturity yields (DGS). The sample period is kept identical---January
	4, 2016 through October 31, 2025---to permit direct comparison with the OIS
	baseline. The purpose is not to argue that DGS is the preferred discount
	benchmark for derivatives, but to test whether the GBM path-risk structure
	survives a change in benchmark curve.
	
	The robustness exercise asks whether the core sign structure survives under
	DGS and which components are most sensitive to benchmark choice. Coefficient
	inference uses date-based HAC (Newey--West) standard errors with maximum lag
	21 trading days.
	
	\subsubsection{In-sample results}
	
	\FloatBarrier
	\begin{table}[H]
		\centering
		\footnotesize
		\caption{In-sample fit summary under the DGS benchmark}
		\label{tab:dgs_is_summary}
		\resizebox{6.5in}{!}{
			\begin{tabular}{lcccccc}
				\toprule
				Specification & Obs. & Trading days & $R^2$ & Adj.\ $R^2$ & RMSE (bp) & MAE (bp) \\
				\midrule
				Common-market + SPX dummy & 48{,}030 & 2{,}457 & 0.229 & 0.229 & 14.96 & 10.56 \\
				SPX separate              & 29{,}377 & 2{,}457 & 0.234 & 0.234 & 14.36 & 9.85 \\
				RUT separate              & 18{,}653 & 2{,}456 & 0.208 & 0.208 & 15.66 & 11.60 \\
				\bottomrule
			\end{tabular}
		}
	\end{table}
	\FloatBarrier
	
	Table~\ref{tab:dgs_is_summary} shows that replacing OIS with DGS weakens fit
	but does not eliminate the structure. The common-market $R^2$ falls from 0.309
	under OIS to 0.229 under DGS, and the market-specific fits decline similarly.
	
	\FloatBarrier
	\begin{table}[H]
		\centering
		\footnotesize
		\caption{In-sample coefficient estimates under the DGS benchmark with HAC(21) inference}
		\label{tab:dgs_is_coef_hac21}
		\begin{tabular}{lccc}
			\toprule
			Regressor
			& Common-market + SPX dummy
			& SPX separate
			& RUT separate \\
			\midrule
			Intercept
			& 29.894*** & 27.172*** & 29.247*** \\
			& (3.638)   & (3.434)   & (3.424)   \\
			
			SPX dummy
			& -0.891    & ---       & ---       \\
			& (0.906)   &           &           \\
			
			$GBM^{DGS,1Y}$
			& -0.471*** & -0.466*** & -0.438*** \\
			& (0.118)   & (0.132)   & (0.110)   \\
			
			$GBM^{DGS,10Y}$
			& 0.273*    & 0.258     & 0.278**   \\
			& (0.140)   & (0.157)   & (0.130)   \\
			
			$BA^{\mathrm{med}}/\tau$
			& 0.152***  & 0.294***  & 0.120***  \\
			& (0.042)   & (0.083)   & (0.031)   \\
			
			$NFCI$
			& -14.456** & -15.285** & -16.532*** \\
			& (5.849)   & (5.937)   & (6.067)    \\
			\bottomrule
		\end{tabular}
		
		\vspace{0.5em}
		\begin{minipage}{0.92\linewidth}
			\footnotesize
			\textit{Notes:} Standard errors in parentheses are date-based HAC
			(Newey--West) standard errors with maximum lag 21 trading days.
			***, **, * denote significance at the 1\%, 5\%, and 10\% levels, respectively.
		\end{minipage}
	\end{table}
	\FloatBarrier
	
	Table~\ref{tab:dgs_is_coef_hac21} shows that the main sign pattern survives:
	$GBM^{DGS,1Y}$ is negative, $GBM^{DGS,10Y}$ is positive,
	$BA^{\mathrm{med}}/\tau$ is positive, and NFCI is negative across all three
	specifications. The short-horizon GBM and bid--ask terms remain strongly
	identified. The weaker component is the long-horizon DGS GBM term, which stays
	positive but is less precisely estimated, especially for SPX. Thus, DGS
	supports sign-pattern robustness while confirming that OIS provides the
	stronger baseline fit.
	
	\subsubsection{Out-of-sample results}
	
	\FloatBarrier
	\begin{table}[H]
		\centering
		\footnotesize
		\caption{LOYO out-of-sample performance summary under the DGS benchmark}
		\label{tab:dgs_oos_summary}
		\resizebox{6.5in}{!}{
			\begin{tabular}{llcccccc}
				\toprule
				Specification & Market & Mean $R^2$ & Median $R^2$ & Pooled $R^2$ & Years with $R^2>0$ & Mean corr. & Mean RMSE (bp) \\
				\midrule
				Common-market & SPX & 0.053  & 0.192 & 0.170 & 7/10 & 0.213 & 15.05 \\
				Common-market & RUT & 0.036  & 0.169 & 0.121 & 6/10 & 0.201 & 16.90 \\
				Separate      & SPX & -0.039 & 0.237 & 0.141 & 7/10 & 0.236 & 15.31 \\
				Separate      & RUT & 0.065  & 0.148 & 0.145 & 7/10 & 0.190 & 16.73 \\
				\bottomrule
			\end{tabular}
		}
	\end{table}
	\FloatBarrier
	
	Table~\ref{tab:dgs_oos_summary} shows modest LOYO performance under DGS. Mean
	$R^2$ is small and turns negative for the SPX separate specification, but
	median and pooled $R^2$ values remain positive throughout, and 6--7 out of 10
	holdout years produce positive $R^2$. As in the OIS case, the LOYO exercise
	should be read as a year-exclusion diagnostic rather than real-time forecasting
	evidence.
	
	The stress pattern differs from the OIS baseline. Under OIS, the 2020 holdout
	is the dominant failure case. Under DGS, the 2022 holdout becomes more
	prominent, especially for SPX, suggesting that the DGS specification is more
	sensitive to the rate-regime transition.
	
	\FloatBarrier
	\begin{table}[H]
		\centering
		\footnotesize
		\caption{LOYO coefficient stability under the DGS benchmark with HAC(21) inference}
		\label{tab:dgs_oos_sign_hac21}
		\resizebox{6.5in}{!}{
			\begin{tabular}{lcccccc}
				\toprule
				Regressor
				& Common sign
				& Common HAC significance
				& SPX sign
				& SPX HAC significance
				& RUT sign
				& RUT HAC significance \\
				\midrule
				$GBM^{DGS,1Y}$
				& $-$\;10/10 & 10/10 at 5\%, 9/10 at 1\%
				& $-$\;10/10 & 9/10 at 5\%, 10/10 at 10\%
				& $-$\;10/10 & 10/10 at 1\% \\
				
				$GBM^{DGS,10Y}$
				& $+$\;10/10 & 4/10 at 5\%, 5/10 at 10\%
				& $+$\;10/10 & 3/10 at 5\%, 4/10 at 10\%
				& $+$\;10/10 & 4/10 at 5\%, 5/10 at 10\% \\
				
				$BA^{\mathrm{med}}/\tau$
				& $+$\;10/10 & 10/10 at 1\%
				& $+$\;10/10 & 10/10 at 5\%, 9/10 at 1\%
				& $+$\;10/10 & 10/10 at 1\% \\
				
				$NFCI$
				& $-$\;10/10 & 7/10 at 5\%, 10/10 at 10\%
				& $-$\;10/10 & 6/10 at 5\%, 9/10 at 10\%
				& $-$\;10/10 & 9/10 at 5\%, 10/10 at 10\% \\
				\bottomrule
			\end{tabular}
		}
		
		\vspace{0.5em}
		\begin{minipage}{0.94\linewidth}
			\footnotesize
			\textit{Notes:} Each entry is computed from the ten leave-one-year-out
			training-sample regressions under the DGS benchmark.
			Standard errors are date-based HAC (Newey--West) standard errors
			with maximum lag 21 trading days.
			The common-market SPX dummy is omitted from the table because it is not part
			of the core GBM block; its sign is mixed across folds.
		\end{minipage}
	\end{table}
	\FloatBarrier
	
	Table~\ref{tab:dgs_oos_sign_hac21} shows that the coefficient signs remain
	stable under DGS. Across the common-market, SPX separate, and RUT separate
	specifications, the short-horizon GBM term is always negative, the long-horizon
	GBM term is always positive, the bid--ask term is always positive, and NFCI is
	always negative.
	
	The HAC significance pattern is more mixed. The short-horizon DGS GBM and
	bid--ask terms remain robust, NFCI remains directionally stable and usually
	significant, but the long-horizon DGS GBM term is weakly identified. Overall,
	the DGS exercise supports the paper's central message with an important
	qualification: the GBM sign structure is not an artifact of the OIS benchmark,
	but OIS delivers stronger fit, sharper long-horizon inference, and greater
	stability around rate-regime transitions.
	
	\section{Discussion}
	\label{sec:discuss}
	
	This section discusses the economic interpretation of the empirical results and
	the limits of the evidence.
	
	\subsection{Economic interpretation of the path-risk structure}
	
	The central empirical result is that the two GBM path-risk terms are repeatedly
	significant with opposite signs. In the main OIS-based specification,
	$GBM^{OIS,1Y}$ loads negatively and $GBM^{OIS,10Y}$ loads positively. The carry
	gap is therefore not a simple discount-rate level effect or a measurement
	residual; it is organized in a way consistent with the path-risk structure of
	parity enforcement.
	
	The opposite signs can be interpreted through the term structure of arbitrage
	capital. A rise in short-term rates mechanically lowers the OIS discount factor,
	which can compress the carry gap, defined as $\log(D^{OIS}/\hat{B})$, when the
	option-implied discount factor adjusts more slowly. By contrast, a higher
	long-horizon yield raises the opportunity cost of allocating scarce capital to
	parity enforcement. Following the capital-allocation logic of \citet{SV97} and
	\citet{BP09}, improved outside opportunities can reduce capital supplied to
	parity enforcement and widen the equilibrium carry gap.
	
	I do not structurally separate these channels. The key point is that the sign
	pattern is stable across in-sample, LOYO, and robustness exercises.
	Residual-stationarity diagnostics reinforce the same conclusion: the fitted and
	observed carry-gap series do not leave a nonstationary residual under the main
	OIS-based specification. Thus, the evidence supports a level relation organized
	by the GBM path-risk structure, while leaving room for regime-dependent
	calibration errors.
	
	The two markets share this structure but differ in channel intensity. SPX and
	RUT both exhibit a positive carry-gap center, low-frequency persistence, and
	links to the GBM path-risk block, trading frictions, and financial conditions.
	The pooled specification is not materially inferior to the separate
	regressions, indicating a common structure. Heterogeneity appears mainly in
	$GBM^{10Y}$ and $BA^{\mathrm{med}}/\tau$: the long-horizon GBM term loads more
	strongly in RUT, while the bid--ask term loads more strongly in SPX.
	
	Put--call parity remains exact as a terminal-payoff identity, and quoted
	price-space residuals are small. The contribution of the carry gap is to make
	the implementation layer visible. A small quoted residual does not imply that
	parity enforcement is economically costless. The positive return to parity
	enforcement is therefore better interpreted as compensation consistent with
	implementation risk, cash-flow timing, and finite capital, rather than as a
	literal free lunch.
	
	\subsection{Benchmark choice, observability, and why now}
	
	The DGS exercise shows that the path-risk structure is not an artifact of the
	OIS curve: under DGS, $GBM^{1Y}$ stays negative, $GBM^{10Y}$ positive,
	$BA^{\mathrm{med}}/\tau$ positive, and NFCI negative. OIS nonetheless fits
	better, with sharper long-horizon inference and greater stability through
	rate-regime transitions. The reason is economic---Treasury yields carry a
	flight-to-quality and convenience-yield component that distorts them precisely
	in the stress episodes when the carry gap is most informative---so OIS remains
	the cleaner baseline and DGS a robustness check.
	
	This also explains why the carry gap is more observable post-GFC, for a reason
	that is institutional before it is statistical. The carry gap is defined
	relative to an OIS discount curve, and OIS discounting is itself a post-crisis
	convention: the multi-curve framework became standard only after 2007--2009, so
	the benchmark leg that defines the gap did not exist in its current form
	earlier. The overnight rate underlying USD OIS---the effective federal funds
	rate before the SOFR transition---was also less tightly anchored under the
	pre-crisis corridor system than under the later floor system, and benchmark
	instability lowers the signal-to-noise ratio of any residual measured against
	it.
	
	These points also bound the claim. Because the pre-GFC benchmark is too noisy to
	support a comparable series, I cannot tell whether the wedge is genuinely larger
	post-GFC or merely better measured; post-crisis balance-sheet and regulatory
	constraints on intermediaries make a structural increase plausible, but the same
	constraints could simply have made an always-present wedge detectable. I
	therefore claim only the weaker proposition---that the post-GFC environment, by
	combining a stable OIS discounting convention, a decade of daily data, the
	\citet{AB21} identification method, and low-cost intraday NBBO data, makes the
	carry gap measurable rather than new.
	
	\subsection{A residual short-end component}
	\label{subsec:unknown_short_end_component}
	
	Explanatory power is weakest at the 1--2 month horizon. This does not negate
	the GBM path-risk term, but suggests that the current specification does not
	exhaust the short end of the carry gap. In pilot tests, an ad hoc
	short-maturity amplification of the GBM term produced only limited improvement,
	suggesting that the issue may reflect a distinct short-end component rather
	than a simple shape misspecification.
	
	The short-maturity residual may reflect microstructure effects, margin timing,
	execution frictions, or other mechanisms that operate most strongly near
	expiration. Beyond documenting the central structure of the carry gap, the
	results therefore point to a short-end-specific component for future work.
	
	\subsection{Limitations and future work}
	
	This study is reduced-form. The coefficients capture conditional associations,
	not structural causal effects. A structural model in which implementation risk
	and capital constraints endogenously generate an equilibrium parity wedge
	remains a natural next step.
	
	The residual-stationarity evidence should also be interpreted narrowly. It
	mitigates the direct spurious-regression concern in the main OIS-based level
	relation, but it does not prove that a time-invariant cointegrating vector
	governs all regimes. Similarly, the LOYO evidence should be read as a
	year-exclusion diagnostic rather than real-time forecasting evidence.
	
	Benchmark choice remains an economic question. DGS confirms that the sign
	structure survives an alternative curve, but also shows that quantitative fit
	and long-horizon inference vary with the benchmark. Future work could examine
	reference curves that more directly reflect the funding costs or opportunity
	costs faced by actual parity enforcers.
	
	The sample is limited to two U.S.\ equity-index option markets. Extending the
	analysis to European or Asian index options, single-name options, or other
	underlying assets would clarify how general the observed structure is. More
	granular data on execution, margin, dealer balance sheets, and market-maker
	capital constraints could also help identify the sources of regime dependence
	and the residual short-end component.
	
	\section{Conclusion}
	\label{sec:conclusion}
	
	This paper separates two propositions about put--call parity: the
	terminal-payoff identity itself and the stronger claim that enforcing it in
	practice is economically risk-free. The former is exact. The latter need not
	follow once daily settlement, variation margin, interim funding needs, and
	finite capital are taken seriously.
	
	The evidence from U.S.\ equity-index options shows that this distinction appears
	not in quoted price space, but in carry space. Futures-based parity residuals
	are compressed near zero, yet the annualized gap between option-implied and
	OIS-benchmark discount factors reveals a systematic carry gap with a positive
	center, low-frequency persistence, and state-variable dependence. The GBM
	path-risk term introduced in this paper provides the central explanatory block:
	its short-to-medium-term and long-term rate-scaled components are repeatedly
	significant, with opposite signs and different magnitudes.
	
	The result is not a claim of precise high-frequency forecastability. LOYO
	validation is used as a year-exclusion diagnostic, and the evidence shows that
	the key coefficient signs remain stable when any individual calendar year is
	excluded. Residual-stationarity diagnostics mitigate the concern that the level
	relation is driven by unrelated persistent trends, and the core sign structure
	also survives replacing OIS with Treasury constant-maturity yields.
	
	The carry gap therefore provides evidence that the economic burden of parity
	enforcement can persist even when visible parity residuals have nearly vanished.
	Markets may compress price-space arbitrage residuals without fully eliminating
	the carry-space wedge created by path-dependent cash-flow timing, funding needs,
	and capital constraints. The post-GFC data environment, the AB21
	within-option-cross-section identification method, and the availability of
	large intraday option datasets make this wedge measurable. The object may not
	be a new economic phenomenon; the contribution is to make it visible, measure
	it, and link it to a reduced-form path-risk structure.
	
	\subsection*{Funding}
	This research did not receive any specific grant from funding agencies in the public, commercial, or not-for-profit sectors.
	
	\subsection*{Declaration of AI usage in manuscript preparation}
	During the preparation of this manuscript, the author used ChatGPT (OpenAI) and Claude (Anthropic) for language refinement and structural clarity.
	All outputs were reviewed and edited by the author, who takes full responsibility for the content.
	
	\subsection*{Declaration of interest}
	The author declares no competing interests.
	
	\newpage
	\onehalfspacing

	\newpage
	\begin{appendices}
		
		\section{Alternative Spot--Futures Robustness}
		\label{app:alt}
		
		This appendix constructs an alternative carry-gap measure from spot--futures
		parity rather than from the option cross-section. The exercise is diagnostic,
		not a replacement for the baseline measure. The main analysis follows
		\citet{AB21}, which identifies option-implied discount factors internally from
		call--put spreads across strikes. The spot--futures route instead combines
		spot, futures, dividend, and OIS inputs. It is therefore noisier by construction,
		but its different error structure makes it useful as an external check on the
		AB21 option-cross-sectional measurement procedure.
		
		\subsection{Futures-implied discount factors}
		
		For an index future with maturity $T$, let $F^{mid}_{t}(T)$ denote the futures
		midpoint, $S_t$ the spot index level, $q_t$ the dividend-yield input, and
		$\tau_t(T)$ the time to maturity in years. The cost-of-carry relation is
		\begin{equation}
			F_t(T)=\frac{S_t\exp(-q_t\tau_t(T))}{B_t(T)}.
		\end{equation}
		Rearranging gives the futures-implied discount factor
		\begin{equation}
			B^{fut}_{t}(T)
			=
			\frac{S_t\exp(-q_t\tau_t(T))}{F^{mid}_{t}(T)},
		\end{equation}
		and the corresponding carry gap
		\begin{equation}
			CG^{fut}_{t}(T)
			=
			\frac{1}{\tau_t(T)}
			\log\left(
			\frac{D^{OIS}_{t}(T)}{B^{fut}_{t}(T)}
			\right).
		\end{equation}
		The empirical analysis uses $10^4 CG^{fut}_{t}(T)$ in basis points.
		
		The construction is implemented at the one-minute frequency. ES futures are
		matched to SPX spot observations, and RTY futures are matched to RUT spot
		observations. The feasible sample is limited by the availability of
		minute-level spot index data, so the common-support samples begin on
		January~3, 2017 for SPX/ES and December~3, 2018 for RUT/RTY. Futures quotes are
		restricted to regular trading hours and filtered for valid bid--ask quotes.
		Dividend and OIS inputs are attached using the most recent available
		observation subject to maximum-lag filters, and the OIS discount factor is
		interpolated to the futures maturity.
		
		I compute two dividend specifications. The first uses the trailing-twelve-month
		dividend yield,
		\begin{equation}
			q_{t,\mathrm{ttm}} = q^{ttm}_t .
		\end{equation}
		The second uses a forward dividend specification based on the rolling growth
		rate of trailing-twelve-month dividends, $Q_{TTM}$:
		\begin{equation}
			q_{t,\mathrm{fwd}}
			=
			q^{ttm}_t
			\frac{\exp(g^Q_t\tau_t)-1}{g^Q_t\tau_t},
		\end{equation}
		with the ratio set to one when $g^Q_t\tau_t$ is numerically zero. For each
		$d\in\{\mathrm{ttm},\mathrm{fwd}\}$,
		\begin{equation}
			B^{fut}_{t,d}(T)
			=
			\frac{S_t\exp(-q_{t,d}\tau_t(T))}{F^{mid}_{t}(T)}
		\end{equation}
		and
		\begin{equation}
			CG^{fut}_{t,d}(T)
			=
			\frac{1}{\tau_t(T)}
			\log\left(
			\frac{D^{OIS}_{t}(T)}{B^{fut}_{t,d}(T)}
			\right).
		\end{equation}
		
		\FloatBarrier
		\begin{table}[H]
			\centering
			\onehalfspacing
			\footnotesize
			\caption{Futures-dominated common support}
			\label{tab:alt_support}
			\begin{threeparttable}
				\begin{tabular}{lcccccc}
					\toprule
					Market & Start date & End date & $\tau_{min}$ & $\tau_{max}$ & Futures rows & Dates \\
					\midrule
					SPX/ES  & 2017-01-03 & 2025-10-31 & 0.0849 & 1.2712 & 7,148 & 2,220 \\
					RUT/RTY & 2018-12-03 & 2025-10-31 & 0.0849 & 1.1507 & 3,244 & 1,732 \\
					\bottomrule
				\end{tabular}
				\begin{tablenotes}
					\small
					\item Notes: The common support is determined by the futures route after imposing
					$\tau\geq 1/12$. The shorter start dates reflect the availability of the
					minute-level spot index series required for the spot--futures match.
				\end{tablenotes}
			\end{threeparttable}
		\end{table}
		\FloatBarrier
		
		\subsection{Comparison design}
		
		I compare three measurement routes: the AB21 option-implied route,
		$CG^{AB21}$, and two spot--futures routes, $CG^{fut}_{\mathrm{ttm}}$ and
		$CG^{fut}_{\mathrm{fwd}}$. The futures route determines the common date and
		maturity support, so the comparison does not mechanically benefit from the
		longer and denser AB21 panel. All observations satisfy $\tau\geq 1/12$.
		
		Each route is estimated using the same OIS-scaled path-risk block as in the
		main text:
		\begin{equation}
			GBM^{OIS,xY}_{i,t}
			=
			10^4\cdot \frac{OIS^{xY}_t}{100}
			\cdot\frac{2}{3}
			\cdot\frac{Vol_{i,t}}{100}
			\cdot\sqrt{\frac{2\tau_{i,t}}{\pi}},
			\qquad x\in\{1,10\},
		\end{equation}
		where $Vol_{i,t}$ is VIX for SPX and RVX for RUT. The pooled full specification is
		\begin{equation}
			\begin{aligned}
				CG^{r}_{i,t}
				=&\ \alpha + \delta D^{SPX}_{i}
				+ \phi_1 GBM^{OIS,1Y}_{i,t}
				+ \phi_{10} GBM^{OIS,10Y}_{i,t} \\
				&+ \beta \mathrm{Fric}^{r}_{i,t}
				+ \gamma NFCI_t
				+ \varepsilon_{i,t},
			\end{aligned}
			\label{eq:alt_route_reg}
		\end{equation}
		where
		\[
		r\in\{AB21,(fut,\mathrm{ttm}),(fut,\mathrm{fwd})\}.
		\]
		Market-specific versions are estimated separately for SPX and RUT. The core
		specification omits the route-specific friction proxy. In the full
		specification, the AB21 route uses the option ATM bid--ask spread divided by
		maturity, whereas the futures routes use the futures relative bid--ask spread
		in basis points divided by maturity. Standard errors are date-based
		Newey--West standard errors with lag 21, with score contributions first
		aggregated by date.
		
		\FloatBarrier
		\begin{table}[H]
			\centering
			\onehalfspacing
			\footnotesize
			\caption{In-sample fit across measurement routes: full specification}
			\label{tab:alt_fit}
			\begin{threeparttable}
				\begin{tabular}{lllrrrrr}
					\toprule
					Route & Panel & Market & Obs. & Days & $R^2$ & RMSE & MAE \\
					\midrule
					$CG^{AB21}$                    & Pooled   & All & 34,033 & 2,206 & 0.302 & 13.87 & 9.28 \\
					$CG^{AB21}$                    & Separate & SPX & 23,690 & 2,206 & 0.330 & 13.24 & 8.69 \\
					$CG^{AB21}$                    & Separate & RUT & 10,343 & 1,726 & 0.278 & 14.79 & 10.46 \\
					$CG^{fut}_{\mathrm{ttm}}$      & Pooled   & All & 10,392 & 2,220 & 0.219 & 30.76 & 22.36 \\
					$CG^{fut}_{\mathrm{ttm}}$      & Separate & SPX &  7,148 & 2,220 & 0.078 & 26.65 & 19.63 \\
					$CG^{fut}_{\mathrm{ttm}}$      & Separate & RUT &  3,244 & 1,732 & 0.051 & 37.72 & 27.46 \\
					$CG^{fut}_{\mathrm{fwd}}$      & Pooled   & All & 10,392 & 2,220 & 0.202 & 29.79 & 21.37 \\
					$CG^{fut}_{\mathrm{fwd}}$      & Separate & SPX &  7,148 & 2,220 & 0.103 & 25.18 & 18.15 \\
					$CG^{fut}_{\mathrm{fwd}}$      & Separate & RUT &  3,244 & 1,732 & 0.041 & 37.13 & 27.05 \\
					\bottomrule
				\end{tabular}
				\begin{tablenotes}
					\small
					\item Notes: RMSE and MAE are in basis points. The full specification includes
					the OIS-scaled 1Y and 10Y GBM terms, the route-specific friction proxy, and NFCI.
				\end{tablenotes}
			\end{threeparttable}
		\end{table}
		\FloatBarrier
		
		Table~\ref{tab:alt_fit} confirms that the futures-implied routes are
		substantially noisier than the AB21 route. Their RMSE and MAE are much larger,
		and their separate-market $R^2$ values are lower. This is expected because the
		futures route combines spot, futures, dividend, and OIS inputs, while the AB21
		route identifies the discount factor internally from the option cross-section.
		The diagnostic question is therefore not whether the futures routes reproduce
		the AB21 route one-for-one, but whether the same path-risk structure remains
		visible under a noisier and independently constructed measure.
		
		\FloatBarrier
		\begin{table}[H]
			\centering
			\onehalfspacing
			\footnotesize
			\caption{Full-specification coefficients across measurement routes}
			\label{tab:alt_coef}
			\begin{threeparttable}
				\begin{tabular}{lllcccc}
					\toprule
					Route & Panel & Market & $GBM^{OIS,1Y}$ & $GBM^{OIS,10Y}$ & Friction & NFCI \\
					\midrule
					$CG^{AB21}$ & Pooled & All & \makecell{-0.617***\\(0.212)} & \makecell{0.465**\\(0.213)} & \makecell{0.152***\\(0.040)} & \makecell{-21.468*\\(11.619)} \\
					$CG^{AB21}$ & Separate & SPX & \makecell{-0.526**\\(0.225)} & \makecell{0.400*\\(0.222)} & \makecell{0.378***\\(0.088)} & \makecell{-28.665**\\(12.064)} \\
					$CG^{AB21}$ & Separate & RUT & \makecell{-0.645***\\(0.170)} & \makecell{0.532***\\(0.184)} & \makecell{0.103***\\(0.023)} & \makecell{-16.521\\(10.115)} \\
					$CG^{fut}_{\mathrm{ttm}}$ & Pooled & All & \makecell{-0.968*\\(0.540)} & \makecell{1.865***\\(0.629)} & \makecell{-0.063\\(0.085)} & \makecell{-11.359\\(23.961)} \\
					$CG^{fut}_{\mathrm{ttm}}$ & Separate & SPX & \makecell{-0.689\\(0.487)} & \makecell{1.645***\\(0.578)} & \makecell{0.011\\(0.083)} & \makecell{-6.246\\(19.179)} \\
					$CG^{fut}_{\mathrm{ttm}}$ & Separate & RUT & \makecell{-1.182*\\(0.678)} & \makecell{1.850**\\(0.824)} & \makecell{-0.083\\(0.101)} & \makecell{-22.512\\(33.482)} \\
					$CG^{fut}_{\mathrm{fwd}}$ & Pooled & All & \makecell{-0.742\\(0.498)} & \makecell{1.617***\\(0.573)} & \makecell{-0.065\\(0.082)} & \makecell{-1.784\\(22.568)} \\
					$CG^{fut}_{\mathrm{fwd}}$ & Separate & SPX & \makecell{-0.318\\(0.436)} & \makecell{1.241**\\(0.513)} & \makecell{0.001\\(0.080)} & \makecell{4.269\\(18.118)} \\
					$CG^{fut}_{\mathrm{fwd}}$ & Separate & RUT & \makecell{-1.138*\\(0.642)} & \makecell{1.788**\\(0.777)} & \makecell{-0.052\\(0.098)} & \makecell{-16.002\\(31.864)} \\
					\bottomrule
				\end{tabular}
				\begin{tablenotes}
					\small
					\item Notes: Date-based Newey--West standard errors with lag 21 are reported in
					parentheses. ***, **, and * denote significance at the 1\%, 5\%, and 10\% levels,
					respectively. The friction proxy is route-specific.
				\end{tablenotes}
			\end{threeparttable}
		\end{table}
		\FloatBarrier
		
		Table~\ref{tab:alt_coef} shows that the long-horizon OIS-scaled path-risk term
		is the most stable component across routes. In the full specification,
		$GBM^{OIS,10Y}$ is positive for $CG^{AB21}$,
		$CG^{fut}_{\mathrm{ttm}}$, and $CG^{fut}_{\mathrm{fwd}}$, and is statistically
		significant in both futures-based routes. The short-horizon GBM term generally
		retains the negative sign, but is weaker in the futures routes, especially for
		SPX. The route-specific friction and NFCI terms are less stable, consistent
		with the noisier and less directly comparable measurement structure of the
		spot--futures route.
		
		I also run leave-one-year-out exercises for each route. As in the main text,
		these exercises are best read as year-exclusion diagnostics rather than
		real-time forecasting tests. The futures route is not a strong level-forecasting
		model, but the sign evidence remains informative: in the pooled core
		specification, $GBM^{OIS,10Y}$ is positive in all nine
		$CG^{fut}_{\mathrm{ttm}}$ folds and in eight of nine
		$CG^{fut}_{\mathrm{fwd}}$ folds. Thus, the futures evidence supports
		sign-stability of the path-risk block rather than precise futures-route
		forecastability.
		
		\subsection{Route-alignment diagnostic}
		
		As a direct comparison of the measured objects, I aggregate all routes to daily
		maturity-bin observations and match AB21 and futures observations by market,
		date, and maturity bin. For each dividend specification
		$d\in\{\mathrm{ttm},\mathrm{fwd}\}$, I estimate
		\begin{equation}
			CG^{fut}_{i,t,b,d}
			=
			a_d + b_d\,CG^{AB21}_{i,t,b} + u_{i,t,b,d},
		\end{equation}
		where $b$ indexes the maturity bin. This exercise does not impose a unit slope.
		The futures route has a different error structure and uses separate spot,
		futures, dividend, and OIS inputs. The diagnostic asks only whether the
		independently constructed object is positively aligned with the AB21 carry gap.
		
		\FloatBarrier
		\begin{table}[H]
			\centering
			\onehalfspacing
			\footnotesize
			\caption{Route alignment: futures-implied carry gap on AB21 carry gap}
			\label{tab:alt_alignment}
			\begin{threeparttable}
					\begin{tabular}{llrrrrrrr}
						\toprule
						Futures route & Market & Obs. & Days & Slope & HAC s.e. & $R^2$ & Corr. & Mean fut.--AB21 \\
						\midrule
						$d=\mathrm{ttm}$ & SPX & 6,953 & 2,205 & 0.475*** & 0.120 & 0.071 & 0.266 & 0.71 \\
						$d=\mathrm{fwd}$ & SPX & 6,953 & 2,205 & 0.320*** & 0.111 & 0.035 & 0.187 & -0.84 \\
						$d=\mathrm{ttm}$ & RUT & 3,219 & 1,720 & 0.873*** & 0.161 & 0.163 & 0.404 & -28.08 \\
						$d=\mathrm{fwd}$ & RUT & 3,219 & 1,720 & 0.790*** & 0.158 & 0.140 & 0.374 & -26.86 \\
						\bottomrule
				\end{tabular}
				\begin{tablenotes}
					\small
					\item Notes: The dependent variable is the futures-implied carry gap
					$CG^{fut}_{i,t,b,d}$ and the regressor is the AB21 carry gap matched by
					market, date, and maturity bin. The final column reports the mean
					futures-implied carry gap minus the mean AB21 carry gap, in basis points.
					Standard errors are date-based Newey--West standard errors with lag 21.
				\end{tablenotes}
			\end{threeparttable}
		\end{table}
		\FloatBarrier
		
		Table~\ref{tab:alt_alignment} shows positive and statistically significant
		alignment slopes in both markets and under both dividend specifications. For
		SPX, the level match is also close: the average futures-minus-AB21 difference
		is 0.71\,bp under the trailing-dividend route and -0.84\,bp under the
		forward-dividend route. For RUT, the futures route has a sizable negative level
		offset of about 27--28\,bp, but the slope and correlation remain positive.
		Thus, the alternative route does not mechanically reproduce the AB21 level in
		all cases, but it preserves a meaningful component of the same low-frequency
		variation.
		
		Overall, the spot--futures route is a conservative external diagnostic, not a
		replacement for the option-implied discount-factor measure. It is noisier and
		has shorter support, but its independent error structure is precisely what
		makes it useful. The repeated positive loading of the long-horizon GBM term and
		the positive AB21--futures alignment make it difficult to attribute the baseline
		results solely to an artifact of the AB21 option-cross-sectional measurement
		procedure.
		
		\newpage
		\section{Time-Series Fit by Maturity Bin}
		\label{app:additional_fig}
		
		This appendix complements the maturity-bin results summarized in
		Section~\ref{sec:is}. The main text reports the cross-bin pattern of fit,
		while this appendix provides the underlying time-series fit across the full
		maturity spectrum. In each figure, the solid line is the actual daily carry gap
		and the gray line is the regression fitted value.
		
		\subsection{Pooled specification}
		
		\FloatBarrier
		\begin{figure}[H]
			\centering
			\includegraphics[width=6.5in]{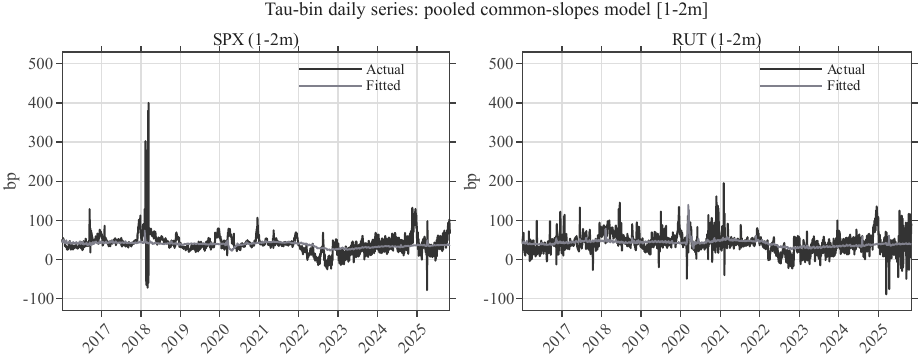}
			\caption{Time-series fit of the pooled specification: 1--2 month maturity bin.}
			\label{fig:app_pool_0102}
		\end{figure}
		\FloatBarrier
		
		\FloatBarrier
		\begin{figure}[H]
			\centering
			\includegraphics[width=6.5in]{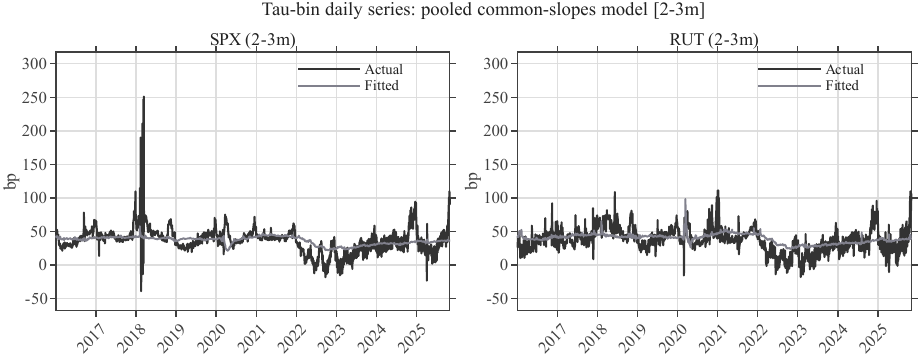}
			\caption{Time-series fit of the pooled specification: 2--3 month maturity bin.}
			\label{fig:app_pool_0203}
		\end{figure}
		\FloatBarrier
		
		\FloatBarrier
		\begin{figure}[H]
			\centering
			\includegraphics[width=6.5in]{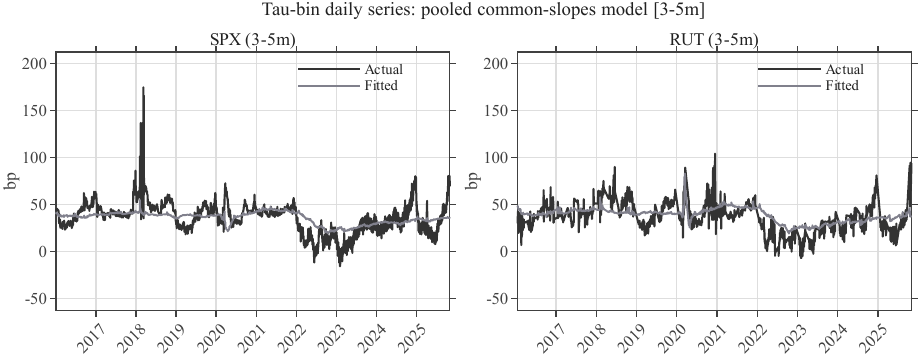}
			\caption{Time-series fit of the pooled specification: 3--5 month maturity bin.}
			\label{fig:app_pool_0305}
		\end{figure}
		\FloatBarrier
		
		\FloatBarrier
		\begin{figure}[H]
			\centering
			\includegraphics[width=6.5in]{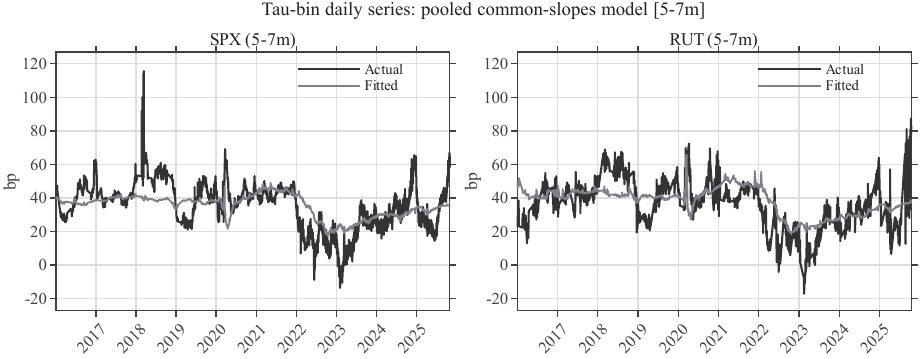}
			\caption{Time-series fit of the pooled specification: 5--7 month maturity bin.}
			\label{fig:app_pool_0507}
		\end{figure}
		\FloatBarrier
		
		\FloatBarrier
		\begin{figure}[H]
			\centering
			\includegraphics[width=6.5in]{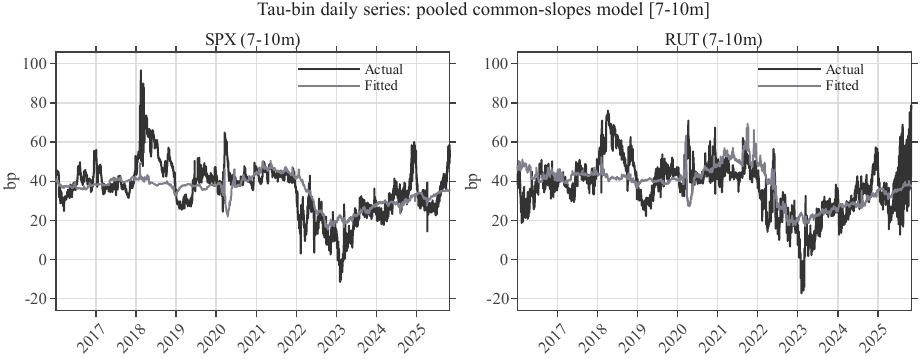}
			\caption{Time-series fit of the pooled specification: 7--10 month maturity bin.}
			\label{fig:app_pool_0710}
		\end{figure}
		\FloatBarrier
		
		\FloatBarrier
		\begin{figure}[H]
			\centering
			\includegraphics[width=6.5in]{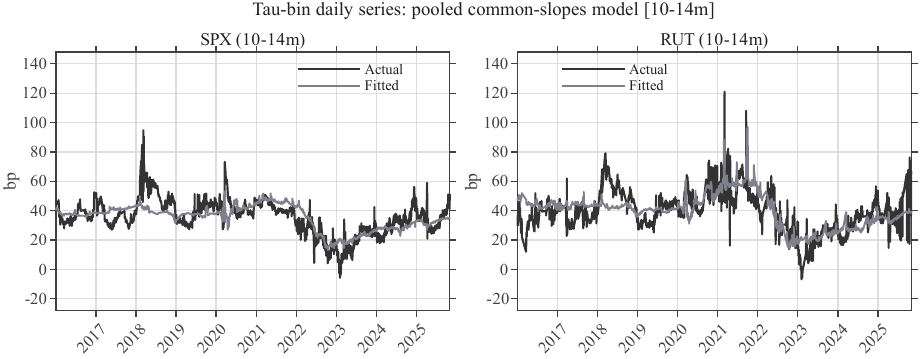}
			\caption{Time-series fit of the pooled specification: 10--14 month maturity bin.}
			\label{fig:app_pool_1014}
		\end{figure}
		\FloatBarrier
		
		\FloatBarrier
		\begin{figure}[H]
			\centering
			\includegraphics[width=6.5in]{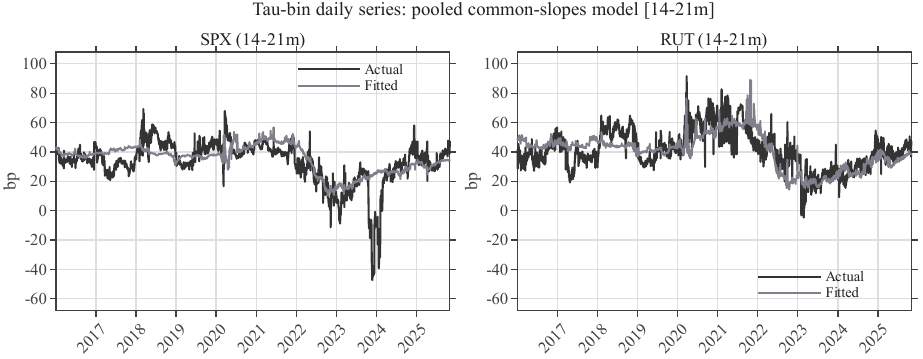}
			\caption{Time-series fit of the pooled specification: 14--21 month maturity bin.}
			\label{fig:app_pool_1421}
		\end{figure}
		\FloatBarrier
		
		\FloatBarrier
		\begin{figure}[H]
			\centering
			\includegraphics[width=6.5in]{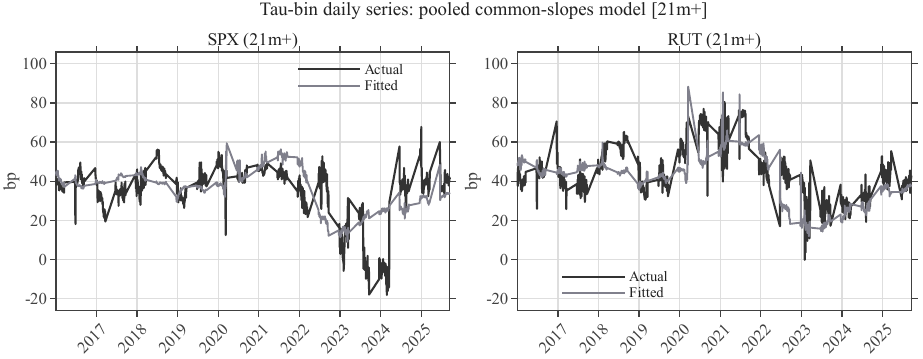}
			\caption{Time-series fit of the pooled specification: beyond 21 months.}
			\label{fig:app_pool_21p}
		\end{figure}
		\FloatBarrier
		
		\newpage
		\subsection{Separate specifications}
		
		\FloatBarrier
		\begin{figure}[H]
			\centering
			\includegraphics[width=6.5in]{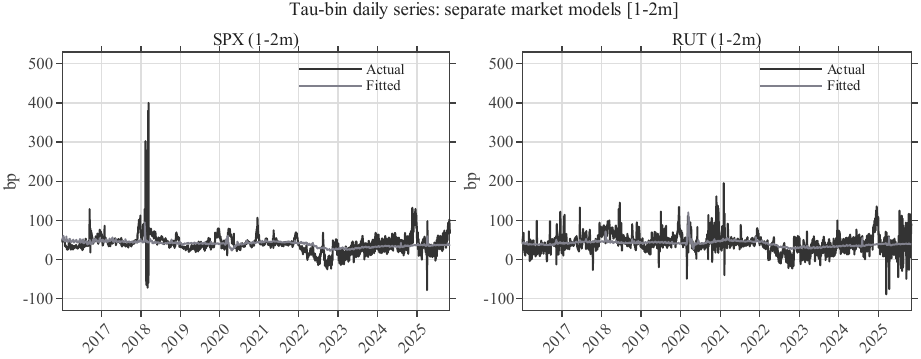}
			\caption{Time-series fit of the separate specification: 1--2 month maturity bin.}
			\label{fig:app_sep_0102}
		\end{figure}
		\FloatBarrier
		
		\FloatBarrier
		\begin{figure}[H]
			\centering
			\includegraphics[width=6.5in]{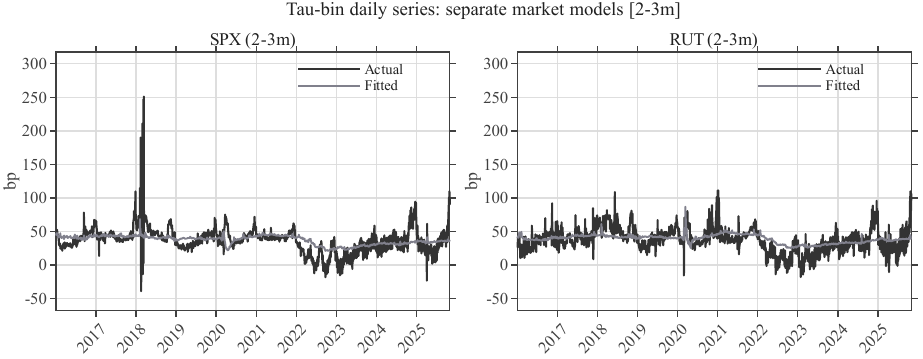}
			\caption{Time-series fit of the separate specification: 2--3 month maturity bin.}
			\label{fig:app_sep_0203}
		\end{figure}
		\FloatBarrier
		
		\FloatBarrier
		\begin{figure}[H]
			\centering
			\includegraphics[width=6.5in]{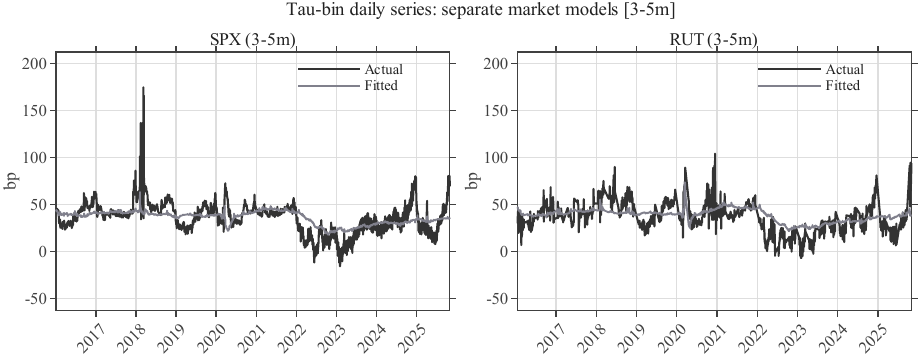}
			\caption{Time-series fit of the separate specification: 3--5 month maturity bin.}
			\label{fig:app_sep_0305}
		\end{figure}
		\FloatBarrier
		
		\FloatBarrier
		\begin{figure}[H]
			\centering
			\includegraphics[width=6.5in]{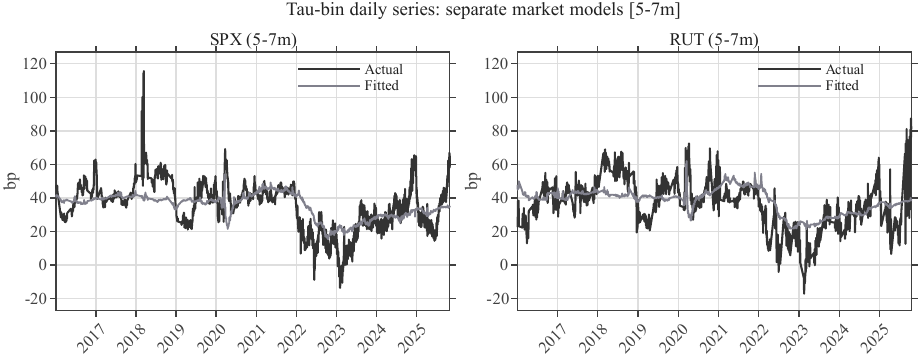}
			\caption{Time-series fit of the separate specification: 5--7 month maturity bin.}
			\label{fig:app_sep_0507}
		\end{figure}
		\FloatBarrier
		
		\FloatBarrier
		\begin{figure}[H]
			\centering
			\includegraphics[width=6.5in]{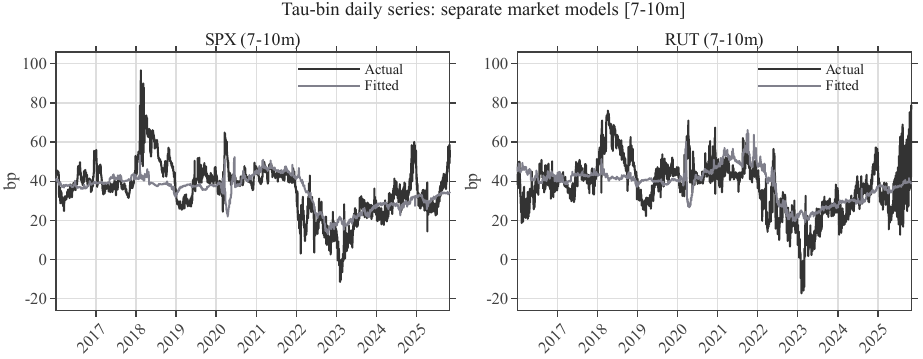}
			\caption{Time-series fit of the separate specification: 7--10 month maturity bin.}
			\label{fig:app_sep_0710}
		\end{figure}
		\FloatBarrier
		
		\FloatBarrier
		\begin{figure}[H]
			\centering
			\includegraphics[width=6.5in]{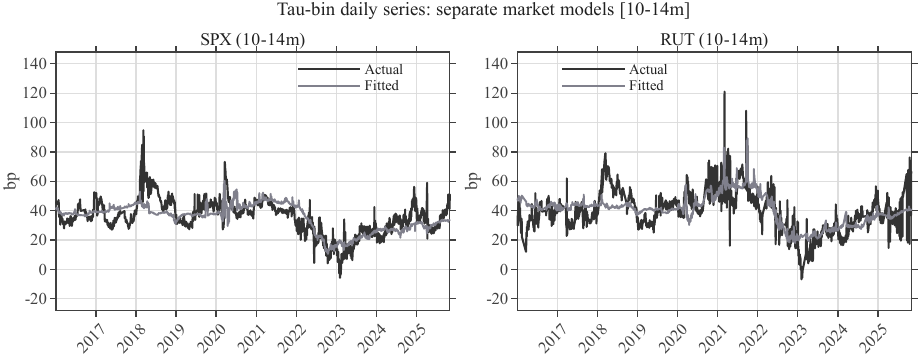}
			\caption{Time-series fit of the separate specification: 10--14 month maturity bin.}
			\label{fig:app_sep_1014}
		\end{figure}
		\FloatBarrier
		
		\FloatBarrier
		\begin{figure}[H]
			\centering
			\includegraphics[width=6.5in]{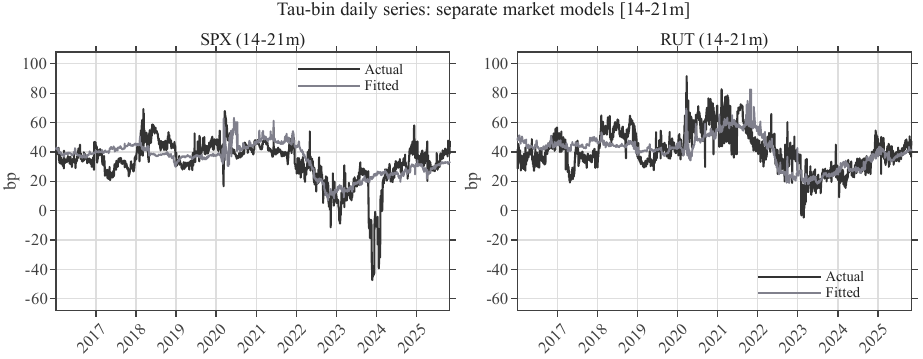}
			\caption{Time-series fit of the separate specification: 14--21 month maturity bin.}
			\label{fig:app_sep_1421}
		\end{figure}
		\FloatBarrier
		
		\FloatBarrier
		\begin{figure}[H]
			\centering
			\includegraphics[width=6.5in]{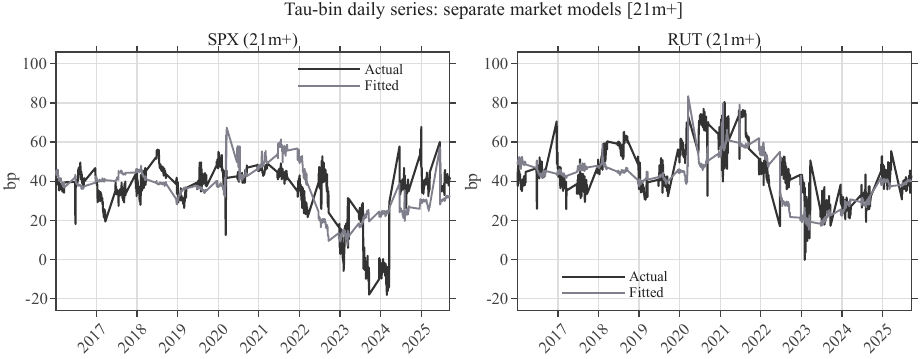}
			\caption{Time-series fit of the separate specification: beyond 21 months.}
			\label{fig:app_sep_21p}
		\end{figure}
		\FloatBarrier
		
	\end{appendices}
	
\end{document}